%% file: main_arxiv.tex
\pdfoutput = 1
\documentclass[letterpaper,12pt]{article} 
\usepackage{fontenc}
\usepackage{epsfig}
\usepackage{rotating}
\usepackage{graphicx}
\usepackage{epstopdf}
\usepackage{setspace}
\usepackage{amsmath}
\usepackage{amssymb}
\usepackage{multirow}
\usepackage{float}
\usepackage{bm}
\usepackage{diagbox}
\usepackage{lscape}
\usepackage{color}
\epsfverbosetrue
\input{defs_letter}

\begin{document}
\pagestyle{empty}
\doublespacing
\input{title}
\newpage
\pagestyle{plain}
\input{sect1}
\input{sect2}
\input{sect3}
\input{sect4}
\input{sect5}

\input{acknowl}
\newpage
\input{append}
\newpage
\onehalfspacing
\input{refs}
\newpage
\input{tabs}
\input{figs_arxiv}
\end{document}

%% file: defs_letter.tex
\def\abstract{\if@twocolumn
\section*{Abstract}
\else \normalsize 
\begin{center}
{\bf Summary\vspace{-.5em}\vspace{0pt}} 
\end{center}
\quotation 
\fi}
\def\endabstract{\if@twocolumn\else\endquotation\fi}

\makeatletter
\@addtoreset{equation}{section}
\makeatother

\newtheorem{remark}{Remark}[section]

\newcommand{\myappendix}[1]{
	\setcounter{section}{1}
        \renewcommand{\thesection}{A\arabic{section}}}

%% file: title.tex
\begin{titlepage}
\title{{\sf Inversion Copulas from Nonlinear State Space Models with Application to Inflation Forecasting}} 
\author{Authors}
\author{{\sf Michael Stanley Smith and Worapree Maneesoonthorn$^\star$}\\ \\
{\sf Melbourne Business School, University of Melbourne}}
\date{}
\maketitle
\vspace{2in}

\noindent
{\small $\mbox{}^\star$ Corresponding author: 
Assistant Professor Ole Maneesoonthorn, Melbourne Business School, 200 Leicester Street, 
Carlton, VIC, 3053, Australia. Email: O.Maneesoonthorn@mbs.edu.}

\newpage
\begin{center}
{\LARGE {\sf Inversion Copulas from Nonlinear State Space Models with Application to Inflation Forecasting}}\\
\vspace{1in}
{\Large {\bf Abstract}}
\end{center}
\vspace{-5pt}
\onehalfspacing
\noindent
We propose to construct copulas from the inversion of nonlinear state space models. These allow for new time 
	series models that have the same serial dependence structure of a state space model, but with an arbitrary marginal distribution, and flexible density forecasts. We examine the time series properties of the copulas, outline serial dependence measures, and estimate the models using likelihood-based methods. Copulas constructed from three example state space 
	models are considered: a stochastic volatility model with an unobserved component, a Markov switching autoregression, and a Gaussian linear unobserved component model. We show that all three inversion copulas with flexible margins improve the fit and density forecasts of quarterly U.S. broad inflation and electricity inflation.
\vspace{2.5in}


\end{titlepage}
\doublespacing

%% file: sect1.tex
\vspace{-15pt}
\section{Introduction}\label{sec:intro}
\vspace{-10pt}
Parametric copulas
constructed through the inversion of a latent multivariate 
distribution (Nelsen~2006, sec. 3.1) 
are popular for the analysis of high-dimensional dependence. For example,
Gaussian (Song~2000), t (Embrechts, McNeil \& Straumann~2001) and
skew t (Demarta \& McNeil~2005; Smith, Gan \& Kohn~2012) distributions
have all been used
to form such `inversion copulas'.
More recently,
Oh \& Patton~(2015) suggest employing distributions formed through
marginalization over
a small number of latent factors.
However, the copulas constructed from these distributions cannot
capture accurately the serial dependence exhibited by many
time series. As an alternative, we instead propose a broad new class of
inversion copulas
formed
by inverting parametric nonlinear state space models. 
Even though the dimension of such a copula
is high, it is parsimonious because its parameters are those
of the underlying latent state space model.
The copula also
has the same serial dependence structure as the state space model. 
But when such copulas are combined with an
arbitrary marginal distribution for the data, they allow for the construction
of new time series models. These models allow for substantially more flexible density forecasts than the underlying state space models themselves, because the latter typically
have rigid margins that are often inconsistent with that observed
empirically.

When the latent state space model is non-stationary, the resulting
copula model for the data is also, but with time invariant
univariate margins.
Alternatively, when the state space model is stationary, so is the 
resulting copula model, and we focus on this case here.
When the state space model is Gaussian and linear,
the resulting inversion copula is a Gaussian copula (Song~2000) with a 
closed form likelihood. 
However,
in general, the likelihood function of a nonlinear state space model cannot be
expressed in closed form. Similarly, neither can the density of the 
corresponding inversion copula. Nevertheless, we show how
existing techniques for computing the likelihood of such state space
models can also be used to compute the copula densities. We also 
provide an efficient spline approximation method for computing the marginal
density and quantile function of the state space model. These are
the most computationally demanding aspects of evaluating
the copula density for time series data. We
outline in detail
how
Bayesian
techniques can be used to compute posterior estimates of the copula model
parameters.
A Markov chain Monte Carlo (MCMC) sampler is used, where
the existing methods for efficiently
sampling the states of a state space model can be employed
directly.
We also study the time series properties of the copula models,
show how to compute
measures of serial dependence, as well as construct density forecasts. We show that the density forecasts from the copula model better reflect the nature of the empirical data distribution than the state space model counterpart in real world applications.

Copula models are used extensively to model cross-sectional dependence, including
between multiple time series; see Patton~(2012) for a review. However,
their use to capture serial dependence is much more limited. Joe~(1997, pp.243-280),
Lambert \& Vandenhende~(2002), Frees \& Wang~(2006), 
Beare~(2010), Smith, Min, Almeida \& Czado~(2010) and 
Loaiza-Maya, Smith \& Maneesoonthorn~(2017) use Archimedean, elliptical
or decomposable vine
copulas to capture serial dependence in univariate time series.
While the likelihood is available in closed form for these copulas, they 
cannot capture as wide a range of serial dependence structures as
the inversion copulas proposed here can.
Moreover,
the proposed inversion copulas are
simple to specify,
often
more parsimonious,
and can be easier to estimate than the copulas used previously. 

Recently, copulas with time-varying parameters have proven popular
for the analysis of multivariate time series data; for example, see
Almeida \& Czado~(2012), Hafner \& Manner~(2012),
De Lira Salvatierra \& Patton~(2015) and Creal \& Tsay~(2015).
However, these authors use
copulas to account for the (conditional) cross-sectional dependence
as in Patton~(2006).
This is completely
different to our objective of constructing a $T$-dimensional
copula for
serial dependence. Semi- and nonparametric
copula functions
(Kauermann, Schellhase \& Ruppert~2013)
can also be used to model serial dependence. However, such an approach
is better suited to longitudinal data, 
where there are repeated observations of the 
time series.

To highlight the broad range of new copulas that can be formed using our approach, we
consider three in detail. 
They are formed by inversion of three
stationary latent state space models that are popular in forecasting macroeconomic time series.
The first is a stochastic volatility
model with an unobserved first order autoregressive mean component. 
The second is
a Markov switching first order autoregression. The
third is a Gaussian unobserved component model, where the 
unobserved component follows a $p$th order autoregression.
When forming an inversion copula,
all characteristics (including moments)
of the marginal distribution of the state space model are lost, 
leaving the parameters potentially
unidentified. For each of the three inversion copulas we study in detail,
we solve this problem by imposing
constraints on the parameter space. We show how 
to implement the MCMC sampling scheme, where the states 
are generated using existing methods,
and the parameters are drawn efficiently from constrained distributions. 
In an empirical setting, we also show how to estimate the copula parameters
using maximum likelihood.

To show that that using an inversion copula with a flexible margin can substantially improve forecast density
accuracy, compared to employing the state space model directly, we use it
to model and forecast quarterly U.S. broad inflation and U.S. electricity inflation. This is a long-standing problem 
on which there is a large literature
(Faust \& Wright~2013).
A wide range of univariate time series models have been used previously,
including the three state space models examined here.
However, all three
have marginal distributions that are inconsistent with that observed empirically for
inflation, which exhibits strong positive skew and heavy tails. Moreover, the predictive
distributions from the state space models are either exactly or approximately symmetric-- a feature
that places excessively high probability on severe deflation.
In comparison, the inversion copula models can employ the same serial dependence
structure as the latent state space models, but also incorporate much more accurate
asymmetric 
marginal distributions. We show that this not only improves the fit of the time
series models, but that it increases the accuracy of the 
one-quarter-ahead density forecasts significantly. 

The rest of the paper is organized as follows. In Section~\ref{sec:icop} we first define a time series copula model, and then outline the construction of
an inversion copula from a nonlinear
state space model. The special case of a Gaussian linear 
state space model is considered separately. 
We then discuss
estimation, time series properties, measures of serial dependence and 
prediction.
In Section~\ref{sec:threecop} we discuss the three inversion
copulas that we examine in detail,
while in Section~\ref{sec:empirical} we
present the analyses of the 
U.S. broad inflation and U.S. electricity inflation;
Section~\ref{sec:disc} concludes. In a supplementary appendix, we provide a simulation study that verifies the proposed methodology in a controlled setting (Section A), as well as supplementary figures for the U.S. electricity inflation application (Section B).

%% file: sect2.tex
\vspace{-15pt}
\section{Time Series Copula Models}\label{sec:icop}
\vspace{-10pt}
Consider a discrete-time stochastic process $\{Y_t\}_{t=1}^T$, 
with a time series
of observed values
$\bm{y}=(y_1,\ldots,y_T)$. Then a copula model
decomposes its joint distribution function as 
\begin{equation}
F_Y(\bm{y})=C(\bm{u})\,. \label{eq:datacdf}
\end{equation}
Here, $\bm{u}=(u_1,\ldots,u_T)$, 
$u_t=G(y_t)$, and $G$ is the marginal distribution function
of $Y_t$,
which we assume to be time invariant. The function $C$ is a $T$-dimensional
copula function (Nelsen~2006, p.45), which captures all serial 
dependence in the data.
All marginal features of the data
are captured by $G$, which can
be modeled separately, and either parametrically or non-parametrically.
While Equation~(\ref{eq:datacdf}) applies equally to both
continuous and discrete-valued time series data, 
we focus here on the former, where
the density
\begin{equation}
f_Y(\bm{y})=\frac{d}{d\bm{y}} F_Y(\bm{y})=c(\bm{u})\prod_{t=1}^T g(y_t)\,.\label{eq:cmod}
\end{equation}
Here, $g(y)=\frac{d}{d y}G(y)$ is the marginal density of each observation,
and $c(\bm{u})=\frac{d}{d \bm{u}}C(\bm{u})$
is widely called the `copula density'.

We refer to Equations~(\ref{eq:datacdf}) and~(\ref{eq:cmod}) as a `time series
copula model'. The main challenge in constructing such a model is the selection
of an appropriate copula function $C$. It has to be both of high dimension,
and also account
accurately for the potentially complex
serial dependence structure in $\{Y_t\}_{t=1}^T$.

\vspace{-10pt}
\subsection{State Space Inversion Copula}\label{sec:ssiv}
\vspace{-5pt}
A popular way to construct a high-dimensional
copula $C$ is by transformation from
a latent continuous-valued stochastic process $\{Z_t\}_{t=1}^T$,
with joint
distribution function $F_Z$.
Let $F_{Z_t}$ be the marginal distribution function of $Z_t$. 
Then, by setting
$U_t=F_{Z_t}(Z_t)$, the $T$ observations of the 
stochastic process $\{U_t\}_{t=1}^T$ have distribution
function
\[
C(\bm{u})=F_Z(F_{Z_1}^{-1}(u_1),\ldots,F_{Z_T}^{-1}(u_T))\,,
\]
and density
function
\begin{equation}
c(\bm{u})=\frac{f_Z(\bm{z})}{\prod_{t=1}^T f_{Z_t}(z_t)}\,,
\label{eq:icop}
\end{equation}
where
$z_t=F_{Z_t}^{-1}(u_t)$,
$\bm{z}=(z_1,\ldots,z_T)$, $f_Z(\bm{z})=\frac{d}{d \bm{z}} F_Z(\bm{z})$, and 
$f_{Z_t}(z_t)=\frac{d}{d z_t} F_{Z_t}(z_t)$.
The transformation ensures that
each $U_t$ is marginally uniformly distributed on $[0,1]$, so that
$C$ meets the conditions of a copula function (Nelsen~2006, p.45).

This approach to copula construction is called inversion by
Nelsen~(2006, sec. 3.1), and we
label such a copula an `inversion copula'.
Table~\ref{tab:icd} depicts the transformations between
$Y_t$, $U_t$ and $Z_t$, along with their distribution functions and domains. 
Previous
choices for $F_Z$ include elliptical (especially 
the Gaussian and t), skew t distributions and latent factor models
(Oh \& Patton~2015).
The dependence properties
of the resulting inversion copulas are inherited from those of
$F_Z$, although all location, scale and other marginal 
properties of $F_{Z_t}$ 
are lost in the transformation.
In this paper we propose to construct an inversion copula
from a latent nonlinear state space model for $\{Z_t\}_{t=1}^T$. 
In doing so, we aim to construct
new high-dimensional copulas that inherit the rich range of serial dependence 
structures that state space models allow.

The nonlinear state space model we consider is generically given by
\begin{eqnarray}
Z_t|\bm{X}_t=\bm{x}_t &\sim &H_t(z_t|\bm{x}_t;\bm{\psi}) \label{eq:obsn}\\
\bm{X}_t|\bm{X}_{t-1}=\bm{x}_{t-1} &\sim &K_t(\bm{x}_t|\bm{x}_{t-1};\bm{\psi}).
\label{eq:trans}
\end{eqnarray}
Here, $H_t$ is the distribution function of
$Z_t$, conditional on a $r$-dimensional state vector $\bm{X}_t$.
The states follow a Markov process, with
conditional distribution function $K_t$.
Parametric distributions are almost always adopted for $H_t$ and $K_t$,
and we do so here with parameters we denote collectively as
$\bm{\psi}$. 
In the time series
literature
Equations~(\ref{eq:obsn}) and~(\ref{eq:trans}) are
called the measurement and transition distributions,
although in the copula context $Z_t$ is not directly
observed, but is also latent.
Note that even though the state vector $\bm{X}_t$ has Markov order one,
$Z_t$ is Markov with an arbitrary order; see Durbin \& Koopman~(2012)
for properties of this model.

A key requirement in evaluating the inversion copula density at
Equation~(\ref{eq:icop}) is computing the marginal
distribution and density functions of $Z_t$. These are given
by
\begin{eqnarray}
F_{Z_t}(z_t|\bm{\psi}) &= &\int H_t(z_t|\bm{x}_t;\bm{\psi})f(\bm{x}_t|\bm{\psi})
\mbox{d}\bm{x}_t \nonumber \\
f_{Z_t}(z_t|\bm{\psi}) &= &\int h_t(z_t|\bm{x}_t;\bm{\psi})f(\bm{x}_t|\bm{\psi})
\mbox{d}\bm{x}_t\,,\label{eq:mdist}
\end{eqnarray}
where the dependence on $\bm{\psi}$ is denoted explicitly here. The density
$h_t(z_t|\bm{x}_t;\bm{\psi}) =\frac{d}{d z_t} H_t(z_t|\bm{x}_t;\bm{\psi})$, 
and $f(\bm{x}_t|\bm{\psi})$ is the marginal density of the state
variable $\bm{X}_t$
which can be derived analytically from the transition
distribution
for most state space models used in practice.
Evaluation of the integrals in 
Equation~(\ref{eq:mdist}) is typically straightforward either 
analytically or numerically, as
we show later 
for three different 
state space models. Note that $z_t$ is a function of $\bm{\psi}$
through the quantile function $z_t=F^{-1}_{Z_t}(u_t|\bm{\psi})$,
as we discuss later.

A more challenging problem is the evaluation of
the numerator in Equation~(\ref{eq:icop}). To compute this,
the state vector $\bm{x}=(\bm{x}_1,\ldots,\bm{x}_T)$ needs to be integrated
out, with
\begin{eqnarray*}
f_Z(\bm{z}|\bm{\psi}) &= &\int f(\bm{z}|\bm{x},\bm{\psi}) f(\bm{x}|\bm{\psi})
\mbox{d}\bm{x} \\
&= &\int \prod_{t=1}^T h_t(z_t|\bm{x}_t;\bm{\psi})
\prod_{t=2}^T k_t(\bm{x}_t|\bm{x}_{t-1};\bm{\psi}) f(\bm{x}_1;\bm{\psi}) \mbox{d}\bm{x}
\,,
\end{eqnarray*}
where $k_t(\bm{x}_t|\bm{x}_{t-1};\bm{\psi})=\frac{d}{d \bm{x}_t}
K_t(\bm{x}_t|\bm{x}_{t-1};\bm{\psi})$.
There are many methods suggested
for evaluating this high-dimensional integral; 
for example, see Shephard \& Pitt~(1997),
Stroud, M\"uller \& Polson~(2003), Godsill, Doucet \& West~(2004), 
Jungbacker \& Koopman~(2007),
Richard \& Zhang~(2007),
Scharth \& Kohn~(2016) and references therein.
In this paper we show how popular Markov chain Monte Carlo (MCMC) methods
for solving this problem can also
be used to estimate 
the inversion copula.

Because all features of the marginal distribution of $Z_t$--- including the marginal
moments--- are
lost when forming the copula, parameters in a state space 
model that uniquely affect
these are unidentified and can be 
excluded from $\bm{\psi}$. 
Moreover, where possible we also impose constraints on the parameters so
that $F_{Z_t}$
has zero mean and unit variance. 
While this has no effect on the copula function $C$, it
aids identification of the parameters in the likelihood.

\vspace{-10pt}
\subsection{Gaussian Linear State Space Inversion Copula}\label{sec:gssiv}
\vspace{-5pt}
The stationary Gaussian linear state space model
encompasses many popular time series 
models;
see 
Ljung~(1999, Sec.~4.3) and
Durbin \& Koopman~(2012, Part~1) for overviews. In this special case, we show
here that the inversion copula is the popular Gaussian copula. 
This state space model is given by 
\begin{eqnarray}
Z_t|\bm{X}_t=\bm{x}_t &\sim &N(\bm{b}\bm{x}_t',\sigma^2) \nonumber \\
\bm{X}_t|\bm{X}_{t-1}=\bm{x}_{t-1} &\sim &N(\bm{x}_{t-1}R',F Q F'), \label{eq:lgssm} 
\end{eqnarray}
where $\bm{X}_t$ is a $(1\times r)$ state vector, 
$\bm{b}$ is a $(1\times r)$ vector, 
$\sigma^2$ is the disturbance variance,
$R$ is a $(r \times r)$ matrix of autoregressive coefficients,
with absolute values of all
eigenvalues less than one. The matrices $F$ and $Q$ are
of sizes $(r \times q)$ and $(q \times q)$, respectively, where $q$ represents the dimension of random components driving the $r$-dimensional state. 
The 
copula parameters are $\bm{\psi}\subseteq\{\bm{b},R,F,Q,\sigma^2\}$, 
depending on the specific state space model adopted. 

To identify the parameters in the likelihood, 
they are constrained so that $E(Z_t)=0$ and $\mbox{Var}(Z_t)=1$.
For the latter, if $\mbox{Var}(\bm{X}_t)=\Sigma_X$, then
$\mbox{Var}(Z_t)=\bm{b}\Sigma_X \bm{b}'+\sigma^2$, where 
$\mbox{vec}(\Sigma_X) = (I_{r^2}-R \otimes R )^{-1} \mbox{vec}(F Q F')$.
This results in the equality constraint $\sigma^2=1-\bm{b}\Sigma_X \bm{b}'$,
along with
nonlinear inequality constraints on the other elements of 
$\bm{\psi}$; something we illustrate further 
for a specific Gaussian state space
model in Section~\ref{sec:ucar4}.

With these constraints, the
margins of $Z_t$ are standard normal. 
This greatly simplifies
evaluation
and estimation of the copula
compared to the general case,
where $f_{Z_t}$ and 
$F_{Z_t}$ in Equation~(\ref{eq:mdist}) depend on $\bm{\psi}$ and need to 
be recomputed whenever $\bm{\psi}$ changes.
Moreover,
$(Z_1,\ldots,Z_T)\sim N(\bm{0},\Omega_\psi)$, with
\begin{eqnarray}
\Omega_\psi = \begin{bmatrix} 
1	&	a_1	& \dots	& a_{T-1}\\  
a_1	&	1	&	\dots	&	a_{T-2}\\  
\vdots	&	\vdots	&	\ddots	&	\vdots\\
 a_{T-1}	& 	a_{T-2} &	\dots	& 1 
 \end{bmatrix}, \nonumber
\end{eqnarray} 
where $a_l=\bm{b} \Gamma(l) \bm{b}'$,
and $\Gamma(l)=\mbox{Cov}(\bm{X}_t,\bm{X}_{t-l})$ denotes the
$l^{\mbox{\scriptsize th}}$ autocovariance matrix of the state vectors,
which can be
computed using the multivariate Yule-Walker equations
(Lutkepohl~2006, pp.26--30). 

Because this is a Gaussian copula, its density is known
explicitly (Song~2000) as
\[
c_{Ga}(\bm{u};\bm{\psi})=|\Omega_\psi|^{-1/2}\exp\left( -\frac{1}{2}\bm{z}
(\Omega_\psi^{-1} - I)\bm{z}'\right)\,.
\]
The copula parameters $\bm{\psi}$ can therefore be estimated using the full
likelihood in Equation~(\ref{eq:cmod}). However, $\Omega_\psi^{-1}$ is usually
computationally demanding to evaluate, so that  $c_{Ga}(\bm{u};\bm{\psi})$
is also. 
Therefore, we instead employ Bayesian methods where the 
latent states
are generated explicitly as part of a MCMC scheme, as we now discuss. This can be 
just as fast and efficient for the inversion copula, as it is for the 
underlying linear Gaussian state space model using the Kalman Filter.

%

\vspace{-10pt}
\subsection{Estimation}\label{sec:copmod}
\vspace{-5pt}
Assuming a parametric margin $G(y_t;\bm{\theta})$
with parameters $\bm{\theta}$,
the likelihood of the time series copula model is
\begin{equation}
f_Y(\bm{y}|\bm{\psi},\bm{\theta})=c(\bm{u};\bm{\psi})\prod_{t=1}^T g(y_t;\bm{\theta})=
f_Z(\bm{z}|\bm{\psi})\prod_{t=1}^T \frac{g(y_t;\bm{\theta})}{f_{Z_t}(z_t|\bm{\psi})}\,,\label{eq:like}
\end{equation}
where the reliance of the copula density on $\bm{\psi}$ is made explicit, and
$z_t=F_{Z_t}^{-1}(G(y_t;\bm{\theta})|\bm{\psi})$.
Given parameters
$(\bm{\psi},\bm{\theta)}$, marginal $G$, and a method to compute $f_{Z_t}$, $F_{Z_t}$
and $F_{Z_t}^{-1}$, evaluation of the likelihood boils down to
evaluation of $f_Z(\bm{z}|\bm{\psi})$. 
There are a range of methods for doing this, usually tailored 
to specific state space models. They include methods based on sequential
importance sampling
(Shephard \& Pitt~1997; Jungbacker \& Koopman~2007; Richard \& Zhang~2007) from
which maximum likelihood estimates (MLEs) can be computed. For example, 
in Section~\ref{sec:MLE},
we compute the MLEs for three inversion copulas outlined in Section~\ref{sec:threecop} for the U.S. broad inflation application.

However, it
is popular to estimate nonlinear state space models using robust 
MCMC methods,
and we focus
on this approach here.
Conditional on the states, the
likelihood is
\begin{equation}
f(\bm{y}|\bm{x},\bm{\psi},\bm{\theta})=
f_Z(\bm{z}|\bm{x},\bm{\psi})\prod_{t=1}^T \frac{g(y_t;\bm{\theta})}{f_{Z_t}(z_t|\bm{\psi})}=
\prod_{t=1}^T
h_t(z_t|\bm{x}_t;\bm{\psi})\frac{g(y_t;\bm{\theta})}{f_{Z_t}(z_t|\bm{\psi})}
\,.
\label{eq:clike}
\end{equation}
Adopting independent priors $\pi_\psi(\bm{\psi})$
and $\pi_\theta(\bm{\theta})$, estimation
and inference from the model can be based on the sampler below.

\noindent {\underline{Sampling Scheme}}

\noindent
Step 1. Generate from 
$f(\bm{x}|\bm{\psi},\bm{\theta},\bm{y})=f(\bm{x}|\bm{\psi},\bm{\theta},\bm{z})\propto \prod_{t=1}^T h_t(z_t|\bm{x}_t;
\bm{\psi})f(\bm{x}|\bm{\psi})$.

\noindent
Step 2. Generate from $f(\bm{\psi}|\bm{x},\bm{\theta},\bm{y}) \propto
\left( \prod_{t=1}^T h_t(z_t|\bm{x}_t;\bm{\psi})/f_{Z_t}(z_t|\bm{\psi}) \right) 
f(\bm{x}|\bm{\psi})
\pi_{\psi}(\bm{\psi})$.

\noindent
Step 3. Generate from $f(\bm{\theta}|\bm{x},\bm{\psi},\bm{y}) \propto
\left( \prod_{t=1}^T
h_t(z_t|\bm{x}_t;\bm{\psi})g(y_t;\bm{\theta})/f_{Z_t}(z_t|\bm{\psi}) \right)
\pi_\theta(\bm{\theta})$.

Note that the values 
$\bm{z}=(z_1,\ldots,z_T)$ are not generated directly in the 
sampling scheme, but instead
are computed for each draw of the parameters $\bm{\psi},\bm{\theta}$.
Crucially,
Step~1 is exactly
the same as that for the underlying state space model, so that the wide range of 
existing procedures for generating the latent states can be employed. 
Step~2 can be undertaken using
Metropolis-Hastings, with a proposal based
on a numerical or other approximation to the conditional posterior. 
However, for some state space
models
it can be more computationally efficient
to generate sub-vectors
of $\bm{\psi}$
from their conditional posteriors, using separate steps. In all our empirical applications in Section~\ref{sec:empirical}, we sample each element of $
\bm{\psi}$ individually using normal approximations
to the each of the conditional posteriors, obtained based on 15 Newton-Raphson steps. 
Initial values for the optimization are the
parameter means obtained from the density
$q(\bm{\psi})\propto f(\bm{x}|\bm{\psi})\pi_\psi(\bm{\psi})$. Note that the prior $\pi_\psi(\bm{\psi})$ reflects the parameter constraints required to identify the inversion copula, resulting in appropriately truncated normal approximations. To speed up the optimization, $F_{Z_1}$ (and its inverse and derivative $f_{Z_1}$) are not updated at each Newton-Raphson step. Nevertheless, we show that 
the resulting normal distributions are appropriate proposal densities
later in our empirical work.

For non-parametric marginal models it is often
attractive to follow Shih \& Louis~(1995) and others, and
employ two-stage estimation, so that Step~3 is not required.
However, for parametric models, $\bm{\theta}$ can be generated
at Step~3 using a Metropolis-Hastings step. 
It is appealing 
to use the posterior from the marginal model as a proposal, with 
density $q(\bm{\theta})\propto \prod_{t=1}^T g(y_t;\bm{\theta})
 \pi_\theta(\bm{\theta})$. However, this should be avoided because
when there is strong dependence --- precisely
the circumstance where the copula model is most useful --- this proposal
can be a poor approximation
to the conditional posterior.

The computations associated with Step 1 is equivalent to those 
of the conventional state space model on which the inversion copula is based.
We outline later key aspects of the samplers
that are used to estimate 
the three specific inversion copulas considered.
However, note that at Steps~2 and~3 the value of $\bm{z}$ needs updating, 
although not at the end of Step~1. When
updating
$z_t=F_{Z_t}^{-1}(G(y_t;\bm{\theta})|\bm{\psi})$, repeated 
evaluation of the quantile function $F_{Z_t}^{-1}$
is the most computationally demanding aspect 
of the sampling scheme.
In
the Appendix we outline how to achieve this quickly and accurately
for a stationary nonlinear state space model 
using spline interpolation. 

\vspace{-10pt}
\subsection{Time Series Properties}\label{sec:tsp}
\vspace{-5pt}
Inversion copulas
at Equation~(\ref{eq:icop}) 
can be constructed from either stationary or non-stationary
state space models for $\{Z_t\}_{t=1}^T$.
Here, stationarity
refers to strong or strict stationarity, rather than weak or covariance
stationarity; eg. see Brockwell \& Davis~(1991, p.12).
When a non-stationary state space model
is used, the copula model at Equation~(\ref{eq:cmod})
is a non-stationary time series model for $\{Y_t\}_{t=1}^T$,
but with a time invariant univariate marginal  
$G$. 
Conversely, when
a stationary latent state space model is employed, it is straightforward to show
that $\{Y_t\}_{t=1}^T$ is also 
stationary. Moreover, $\{Z_t\}_{t=1}^T$ and $\{Y_t\}_{t=1}^T$
share the same Markov order.
For the rest of the paper we
only consider the stationary case with Markov order $p$. In which case
$F_{Z_t}$ is time invariant, so that we denote it
simply as $F_{Z_1}$ throughout.

An alternative representation of the copula density at Equation~(\ref{eq:icop})
can be derived as follows. For $a<b$, we employ the
notation
$\bm{z}_{a:b}=(z_a,z_{a+1},\ldots,z_b)$, with
analogous definitions for $\bm{x}_{a:b}$
and $\bm{u}_{a:b}$. Then, 
the (time invariant)
$r$-dimensional marginal density of the latent process for $r\geq 2$ is
\begin{equation*}
f_Z^{(r)}(\bm{z}_{t-r+1:t})
=\int \cdots \int \left( \prod_{s=t-r+1}^t h_s(z_s|\bm{x}_s;\bm{\psi}) \right) 
f(\bm{x}_{t-r+1:t}|\bm{\psi}) \mbox{d}\bm{x}_{t-r+1:t}\,,
\end{equation*}
where $t\geq r$ and $f(\bm{x}_{t-r+1:t}|\bm{\psi})$ is the $r$-dimensional marginal
density of the states $\bm{X}_{t-r+1:t}$.
Then the $r$-dimensional marginal
copula density can be defined as
\begin{equation*}
c^{(r)}(\bm{u}_{t-r+1:t}) =
\frac{f_Z^{(r)}(\bm{z}_{t-r+1:t})}{\prod_{s=t-r+1}^t f_{Z_1}(z_s)}\,.
\end{equation*}
and the
density at Equation~(\ref{eq:icop}) can then be written
in terms of $\{c^{(r)};r=2,\ldots,p+1\}$ as 
\begin{eqnarray}
c(\bm{u}) &= &\prod_{t=p+1}^T f(u_t|\bm{u}_{t-p:t-1})\prod_{t=2}^p f(u_t|\bm{u}_{1:t-1})  
\nonumber \\
 &= &\prod_{t=p+1}^T \frac{c^{(p+1)}(\bm{u}_{t-p:t})}{c^{(p)}(\bm{u}_{t-p:t-1})}
\prod_{t=2}^p \frac{c^{(t)}(\bm{u}_{1:t})}{c^{(t-1)}(\bm{u}_{1:t-1})} \,,
\label{eq:cexp2}
\end{eqnarray}
where we define $c^{(r)}=1$ whenever $r\leq 1$,
and a product to be equal
to unity whenever its upper limit is less than its lower limit. 
For example, for a 
Markov order $p=1$ series, $c(\bm{u})=\prod_{t=2}^T c^{(2)}(\bm{u}_{t-1:t})$,
so that
the marginal copula $c^{(2)}$ fully captures the 
serial dependence structure. In Appendix~B, we show how to
construct $c^{(2)}$ for the
copulas in Section~\ref{sec:threecop}.

\vspace{-10pt}
\subsection{Serial Dependence and Prediction}\label{sec:serial}
\vspace{-5pt}
Measures of serial dependence at a given lag $l\geq 1$, can be computed
from the inversion copula. 
They include
Kendall's tau, Spearman's rho, and
measures of quantile dependence; see Nelsen~(2006, Ch.~5) for an introduction
to such measures of concordance. These can be computed
from the bivariate copula of the observations of the series at times $s$ and 
$t=s+l$ as follows. If the density and distribution 
functions of $(Z_s,Z_t)$ are denoted as
\begin{eqnarray}
f_Z^{s,t}(z_s,z_t|\bm{\psi}) &= &\int \int h_s(z_s|\bm{x}_s;\bm{\psi})h_t(z_t|\bm{x}_t;\bm{\psi})
f(\bm{x}_s,\bm{x}_t|\bm{\psi})\mbox{d}\bm{x}_s\mbox{d}\bm{x}_t\,, \nonumber\\
F_Z^{s,t}(z_s,z_t|\bm{\psi}) &= &\int \int H_s(z_s|\bm{x}_s;\bm{\psi})H_t(z_t|\bm{x}_t;\bm{\psi})
f(\bm{x}_s,\bm{x}_t|\bm{\psi})\mbox{d}\bm{x}_s\mbox{d}\bm{x}_t\,,\label{eq:stdens}
\end{eqnarray}
then the bivariate copula function is
\[
C^{s,t}(u_s,u_t|\bm{\psi})=F_Z^{s,t}(F_{Z_1}^{-1}(u_s|\bm{\psi}),F_{Z_1}^{-1}(u_t|\bm{\psi})|\bm{\psi})\,,
\]
with corresponding
density $c^{s,t}(u_s,u_t|\bm{\psi})=f_Z^{s,t}(z_s,z_t|\bm{\psi})/f_{Z_1}(z_s|\bm{\psi})f_{Z_1}(z_t|\bm{\psi})$. Kendall's tau,
Spearman's rho and the lower quantile dependence for quantile $0<\alpha<0.5$, are then 
\begin{eqnarray*}
\tau_l &= &4\int \int C^{s,t}(u,v)c^{s,t}(u,v)\mbox{d}u\mbox{d}v-1 \\
r_l &= &12\int\int uv c^{s,t}(u,v)\mbox{d}u\mbox{d}v-3\\
\lambda^{--}_l(\alpha) &= &\mbox{Pr}(u_t<\alpha |u_s<\alpha) = C^{t,s}(\alpha,\alpha)/\alpha\,.
\end{eqnarray*}
Quantile dependencies in other quadrants, 
$\lambda^{++}_l(\alpha)=\mbox{Pr}(u_t>1-\alpha|u_{s}>1-\alpha)$, 
$\lambda^{+-}_l(\alpha)=\mbox{Pr}(u_t>1-\alpha|u_{s}<\alpha)$ and 
$\lambda^{-+}_l(\alpha)=\mbox{Pr}(u_t<\alpha|u_{s}>1-\alpha)$ are computed similarly.
We note that 
for $l=1$, the marginal copula $c^{(2)}=c^{s+1,s}$.

For the copulas employed in Section~\ref{sec:threecop}, the integrals at
Equation~(\ref{eq:stdens}) and the dependence metrics can
be computed numerically. However, this may be impractical for
some state space models.
In this case, following Oh \& Patton~(2013),
the dependence measures are readily calculated
via simulation. 
To do this,
simply
generate $(z_1,\ldots,z_{l+1})$ from the state space model, and then 
transform each value to $u_t=F_{Z_1}(z_t)$.
If the iterates
$(u_1^{[j]},\ldots,u_{l+1}^{[j]})$, $j=1,\ldots,J$, are generated in this way,
then Spearman's rho for pairwise dependence between $Y_{t+l}$ and $Y_{t}$ is
\[
r_l=12E(u_{l+1}u_{1})-3 \approx \frac{12}{J}\sum_{j=1}^J\left( u_{l+1}^{[j]}
u_1^{[j]}\right) - 3\,.
\]
The same Monte Carlo iterates can be used to approximate the other
measures of dependence. While
large Monte Carlo samples (e.g. $J=50,000$) 
can be required for these estimates 
to be accurate, simulating from
the marginal copula is both fast and can be undertaken in parallel, so that 
it is not a problem in practice.

The forecast density for $Y_{t+l}$ conditional on $\bm{Y}_{1:t}=\bm{y}_{1:t}$, is
\begin{equation}
f(y_{t+l}|\bm{y}_{1:t},\bm{\theta},\bm{\psi})
= f(z_{t+l}|\bm{z}_{1:t},\bm{\psi})\frac{g(y_{t+l}|\bm{\theta})}{f_{Z_1}(z_{t+l}|\bm{\psi})}\,.\label{eq:preddist}
\end{equation}
Here, $f(z_{t+l}|\bm{z}_{1:t},\bm{\psi})$ is the predictive density for 
$Z_{t+l}$ conditional on $\bm{Z}_{1:t}=\bm{z}_{1:t}$, which can be computed either analytically
or numerically for many state space models, including those
in Section~\ref{sec:threecop} below. Otherwise, the predictive density can be
evaluated via simulation--- a process which is both straightforward 
and fast. First, simulate
a ray of values from the predictive distribution of the state space model
$(z_{t+1},\ldots,z_{t+l}) \sim F(z_{t+1},\ldots,z_{t+l}|\bm{z}_{1:t},\bm{\psi},\bm{\theta})$.
Then 
$y_{t+l}=G^{-1}(F_{Z_1}(z_{t+l}|\bm{\psi})|\bm{\theta})$ is an iterate from the predictive 
distribution.
The predictive distribution can be evaluated
conditional on either
point estimates of $(\bm{\psi},\bm{\theta})$, or over
the sample of parameter values from the posterior.
The latter approach integrates out parameter uncertainty
in the usual Bayesian fashion, and is undertaken in all our empirical work.

%% file: sect3.tex
\vspace{-15pt}
\section{Three Inversion Copulas}\label{sec:threecop}
\vspace{-10pt}
We consider three time series inversion copulas in detail.
The first two are constructed from two popular nonlinear state space models
and cannot be expressed in closed form, 
while the last is constructed from a
linear Gaussian state space model. In each case, we
outline constraints required to identify the parameters when forming the 
copula by inversion, 
as well as how to implement the generic sampler in Section~\ref{sec:copmod}. 
We illustrate the effectiveness of the three
copula models
in Section~\ref{sec:empirical}.

\vspace{-10pt}
\subsection{Stochastic Volatility Inversion Copula}\label{sec:sv}
\vspace{-5pt}
The conditional variance of many
financial and economic time series exhibit strong positive serial dependence.
A popular model used to capture this
is the stochastic volatility model, although
a major limitation is that its
marginal distribution
is symmetric, which is inconsistent
with most series.
Our approach allows for the construction of
time series models that have the same 
serial dependence as a stochastic volatility model,
but also an arbitrary margin that can be asymmetric.

We consider the stochastic
volatility model with an unobserved autoregressive component (SVUC)
given by
\begin{eqnarray}
Z_t|\bm{X}_t=\bm{x}_t &\sim &N(\mu_t,\exp(\zeta_t)) \nonumber \\
\mu_t |\bm{X}_{t-1}=\bm{x}_{t-1} &\sim &N(\bar \mu+\rho_{\mu}(\mu_{t-1}-\bar \mu),\sigma_{\mu}^2) 
\nonumber \\
\zeta_t|\bm{X}_{t-1}=\bm{x}_{t-1} &\sim &N(\bar \zeta+\rho_\zeta (\zeta_{t-1}-\bar \zeta),\sigma_\zeta^2)\,, \label{eq:svuc} 
\end{eqnarray}
where $\bm{x}_t=(\mu_t,\zeta_t)$ is the state vector.
We constrain $|\rho_\mu|<1$ and $|\rho_\zeta|<1$, ensuring $\{Z_t\}$
is a
(strongly) stationary first order Markov processes. The marginal
mean $E(Z_t)=\bar \mu$, so that we set $\bar \mu=0$. 
The marginal variance
$\mbox{Var}(Z_t)=s^2_\mu+\exp(\bar \zeta + s^2_\zeta/2)$,
where
$s^2_\mu=\sigma^2_\mu/(1-\rho_\mu^2)$ and $s^2_\zeta=\sigma^2_\zeta/(1-\rho_\zeta^2)$.
Setting this equal to unity provides an equality constraint 
on $\bar \zeta = \log(1-s^2_\mu)-\frac{s_\zeta^2}{2}$. In addition,
$\exp(\bar \zeta+s^2_\zeta/2)\geq 0$, giving the inequality constraint
$0<\sigma^2_\mu\leq (1-\rho_\mu^2)$.
With these constraints, the 
dependence parameters of the resulting inversion copula are
$\bm{\psi}=\{\rho_\mu,\rho_\zeta,\sigma^2_\mu,\sigma^2_\zeta\}$. 

The marginal density at Equation~(\ref{eq:mdist}) is 
\[
f_{Z_1}(z;\bm{\psi})=\int\int \phi_1\left( z ;\mu,\exp(\zeta)\right) 
\phi_1(\zeta;\bar \zeta,s_\zeta^2) \phi_1(\mu;0,s^2_\mu)d\mu d\zeta\,,
\]
where $\phi_1(z;a,b^2)$ is a univariate Gaussian density with mean $a$ and 
variance $b^2$. The inner integral in $\mu$ can be computed analytically
and (with a little algebra) 
the marginal density and distribution functions are
\begin{eqnarray*}
f_{Z_1}(z;\bm{\psi}) &= &\int \phi_1(z;0,w(\zeta)^2)\phi_1(\zeta;\bar \zeta,s_\zeta^2)d\zeta \\
F_{Z_1}(z;\bm{\psi}) &= &\int \Phi_1(z;0,w(\zeta)^2)\phi_1(\zeta;\bar \zeta,s_\zeta^2)d\zeta\,,
\end{eqnarray*}
with $w(\zeta)^2=s^2_\mu+\exp(\zeta)$. Computing the (log) copula
density at Equation~(\ref{eq:icop}) requires
evaluating $\log(f_{Z_1})$ and the quantile function $F_{Z_1}^{-1}$ at all $T$
observations.
Appendix~A outlines 
how to compute these numerically using 
spline interpolation of both functions. 
In our empirical work we find these
spline-based approximations to be accurate within
5 to 9 decimals places, and fast because
they require direct
evaluation of $F_{Z_1}^{-1}$ at only one point.

We label the inversion
copula constructed from the SVUC model as `InvCop1'. 
Appendix~B outlines how to compute the marginal copula density
$c^{(2)}(u_t,u_{t-1}|\bm{\psi})$
for this copula. Because the time series has
Markov order one, this bivariate
copula characterizes the 
full serial dependence structure.
For example, 
Figure~\ref{fig:svcop}(a) plots $c^{(2)}$ for the
case when
there is no
unobserved mean component (ie. $\rho_\mu=\sigma^2_\mu=0$),
$\rho_\zeta=0.952$ and $\sigma_\zeta=0.045$--- typical values
arising when fitting asset return data. The copula density
is far from uniform,
with high equally-valued quantile
dependence in all four quadrants
($\lambda^{++}_1(0.1)=0.1428,
\lambda^{++}_1(0.05)=0.0964$ and $\lambda^{++}_1(0.01)=0.0454$).
This is
a high level of first order serial dependence, yet $\tau_1=r_1=0$.
This is because $\tau_1$ and $r_1$ measure `level' dependence,
whereas
this copula instead captures bivariate dependence in the second moment. 
Most existing parametric copulas are not well-suited to represent
such
serial dependence;
see Loaiza-Maya et al.~(2016) for a discussion.

We employ the prior $\pi_{\psi}(\bm{\psi})\propto 
\frac{1}{\sigma^2_\mu \sigma^2_\zeta}I(\bm{\psi} \in R_\psi)$, 
where $R_\psi$ is the 
region of feasible parameter values conforming to the restrictions listed above, and $I(X)=1$ if $X$ is true, and zero 
otherwise. 
We outline here how to implement the Step 1 of
the sampling scheme in Section~\ref{sec:copmod}. We
partition the state vector into
$\bm{\mu}=(\mu_1,\ldots,\mu_T)$ and $\bm{\zeta}=(\zeta_1,\ldots,\zeta_T)$, and
use the two separate steps:
\begin{itemize}
\item[] Step~1a. Generate from $f(\bm{\mu}|\bm{\psi},\bm{\zeta},\bm{\theta},\bm{y})
\propto \prod_{t=1}^T \phi_1 \left( z_t;\mu_t;\exp(\zeta_t)\right)f(\bm{\mu}|\bm{\psi})$
\item[] Step~1b. Generate from $f(\bm{\zeta}|\bm{\psi},\bm{\mu},\bm{\theta},\bm{y})
\propto \prod_{t=1}^T \phi_1 \left( z_t;\mu_t;\exp(\zeta_t)\right)f(\bm{\zeta}|\bm{\psi})$
\end{itemize}
The posterior of $\bm{\mu}$ in Step~1a can be recognized as 
normal with zero mean and a band 1 precision
matrix, so that generation is both straightforward and fast.
There are a number of efficient methods to generate $\bm{\zeta}$ in Step~1b
in the literature, 
and we
employ the `precision sampler' for the latent 
states outlined in Chan \& Hsiao~(2014). This is a fast sparse
matrix implementation of the auxiliary mixture sampler (Kim, Shepherd
\& Chib~1998) that is
known to mix well for the stochastic volatility model.

\vspace{-10pt}
\subsection{Markov Switching Inversion Copula}\label{sec:rs}
\vspace{-5pt}
Another
popular class of nonlinear state space models are regime switching models,
which allow 
for structural changes in the dynamics of a series. 
In these models latent regime indicators
usually follow an ergodic Markov chain, in which case the model is
called a Markov switching model; see Hamilton~(1994; Ch.22) for an introduction.

We consider a two regime Markov switching first order autoregressive
model (MSAR1) given by
\begin{eqnarray}
Z_t|\bm{X}_t=\bm{x}_t \sim N(c_{s_t}+\rho_{s_t}z_{t-1},\sigma^2_{s_t}) \nonumber \\
\mbox{Pr}(s_t=j|s_{t-1}=i)=p_{ij}, \label{eq:msar} 
\end{eqnarray}
for regimes $s_t\in\{1,2\}$.
This is a nonlinear state space model 
with state vector $\bm{x}_t=\left(z_{t-1},s_t\right)$.
We assume the Markov chain is ergodic, so that the marginal 
distribution of $s_t$ is time invariant with
$\mbox{Pr}(s_t=1)=\pi_{1}=(1-p_{22})/(2-p_{11}-p_{22})$ and
$\mbox{Pr}(s_t=2)=\pi_{2}=1-\pi_1$. 
Denoting
$s_i^2=\sigma_i^2/(1-\rho_i^2)$, it can be shown that 
stationarity results from the constraints
$|\rho_j|<1$ for $j=1,2$, and $s_i^2 s_j^2 -\rho_j^2(s_i^2)^2>0$ for
$(i,j)=(1,2),(2,1)$. These provide inequality constraints on 
each element of $\{\rho_1,\rho_2,\sigma^2_1,\sigma^2_2\}$, given values
for the other elements.
Following standard practice 
we identify the two components by assuming $\pi_1<\pi_2$. 

When forming the copula, we assume the marginal mean
$E(Z_t)=\bar \mu =\sum_{i=1,2}\pi_i\frac{c_i}{1-\rho_i}=0$ and 
variance 
$\mbox{Var}(Z_t)=\sum_{i=1,2}\pi_i\frac{\sigma^2_i}{1-\rho^2_i}=1$.
This provides the additional equality constraints
\begin{eqnarray*}
c_1 &= &-\frac{\pi_2c_2(1-\rho_1)}{\pi_1 (1-\rho_2)}\,,\\
\sigma^2_1 &= &\frac{(1-\rho_1^2)}{\pi_1}\left(1-\pi_2 s^2_2\right)\,.
\end{eqnarray*}
Also, because $\sigma^2_1>0$, from the last equality constraint above
it follows that $\pi_2 s^2_2<1$, which can be satisfied by imposing an
upper bound on $p_{22}$.
The inversion copula parameters are therefore
$\bm{\psi}=\{c_2,\rho_1,\rho_2,\sigma_2^2,p_{11},p_{22}\}$, subject to 
the inequality constraints above. 

The marginal distribution of $Z_t$
is a mixture of two Gaussians, with
\begin{eqnarray}
f_{Z_1}(z|\bm{\psi})=\sum_{i=1,2} \phi_1\left(z;\mu_i,
s_i^2\right)\pi_i,\label{ms_uni}
\end{eqnarray}
where $\mu_i=\frac{c_i}{1-\rho_i}$. Both $f_{Z_1}$ and $F_{Z_1}$ are
therefore fast to compute, and
the quantile function $F_{Z_1}^{-1}$ is computed using the spline 
interpolation method outlined in
Appendix~A.
The marginal
copula $c^{(2)}$ is given
in Appendix~B, and is
the inversion copula of a mixture of four bivariate Gaussians. 
This is very different than the more
common `mixture copula', which
is a finite mixture of copulas;
for example, see Patton~(2006).  

We label the inversion copula constructed from the 
MSAR1 model as `InvCop2'. 
Unlike the other two inversion copulas examined,  
it can exhibit asymmetric
first order serial
dependence.
To illustrate, Figure~\ref{fig:svcop}(b) plots the
marginal copula $c^{(2)}$ when $p_{11}=0.92$, $p_{22}=0.95$, $\sigma^2_2=0.6$,
$\rho_1=-0.5$, $\rho_2=0.6$ and $c_2=0.02$. In this case,
$r_1=0.159$, $\tau_1=0.113$, and quantile dependence is different in each
quadrant
with
$\lambda^{--}_1(0.1)=0.249$, $\lambda^{++}_1(0.1)=0.201$, 
$\lambda^{+-}_1(0.1)=0.141$ and $\lambda^{-+}_1(0.1)=0.144$.

As before,
we employ the MCMC algorithm in Section~\ref{sec:copmod} to estimate the model. 
The prior $\pi_{\psi}(\psi)\propto \frac{1}{\sigma^2_2}I(\bm{\psi}\in
R_\psi)$, 
where $R_\psi$ is the region of feasible parameter values outlined 
above.
A forward filtering and backward sampling algorithm (Hamilton~1994, p.694)
is used 
to sample the 
regime indicators $\bm{s}=(s_1,...,s_T)$ in Step~1.

\vspace{-10pt}
\subsection{Gaussian Unobserved Component Inversion Copula}\label{sec:ucar4}
\vspace{-5pt}
We also construct an inversion copula from a Gaussian
unobserved component model, where the component
follows a stationary order $p$ autoregression, so that
\begin{eqnarray}
Z_t|\bm{X}_t = \bm{x}_t &\sim &N(\mu_t,\sigma^2) \nonumber \\
\mu_t &= &\bar \mu + \sum_{j=1}^p \rho_j(\mu_{t-j}-\bar{\mu})+\sigma^2_\mu\, \label{eq:ucar}.
\end{eqnarray}
This model (labeled here as UCAR$p$) can be written 
in state space form at Equation~(\ref{eq:lgssm}) with state
vector $\bm{x}_t=(\mu_t,\mu_{t-1},\ldots,\mu_{t-p+1})$ and appropriate
choices for matrices $\bm{b},R,F$ and $Q$. The resulting inversion
copula is a
Gaussian copula with the specific time series dependence structure,
and is labeled `InvCop3'.

We follow Barndorff-Neilsen \& Schou~(1973) and others, and
re-parametrize the autoregressive coefficients by
the partial correlations $\bm{\pi}=(\pi_1,\pi_2,\ldots,\pi_p)$
via the Durbin-Levinson algorithm. An advantage is that stationarity is easily
imposed by the inequalities
$|\pi_j|<1$ for $j=1,\ldots,p$.
When forming the copula, the marginal mean $E(Z_t)=\bar \mu=0$. A second
equality constraint $\sigma^2=1-\mbox{Var}(\mu_t)$, where
$\mbox{Var}(\mu_t)=\sigma^2_\mu \prod_{j=1}^p(1-\pi_j^2)^{-1}$, ensures
the marginal variance is unity.
The parameters of this copula are therefore
$\bm{\psi}=\{\bm{\pi},\sigma^2_\mu\}$, and the prior is $\pi_\psi(\bm{\psi})
\propto \frac{1}{\sigma^2_\mu}I(|\pi_j|<1)$. 

As in Section~\ref{sec:sv},
we use the precision sampler to sample the latent states
$(\mu_1,\ldots,\mu_T)$ at Step~1.
In Step~2 of the scheme, the partial correlations $\bm{\pi}$
are sampled jointly using Metropolis-Hasting with a multivariate
normal approximation proposal computed as described in Section~\ref{sec:copmod}, and
truncated
to the unit cube. The parameter
$\sigma_\mu^2$ is also generated using a truncated
normal approximation as a proposal. We show these are adequate proposals
in our empirical work. 

%% file: sect4.tex
\vspace{-15pt}
\section{Empirical Analysis}\label{sec:empirical}
\vspace{-10pt}
The three inversion copulas in Section~\ref{sec:threecop} are used to model quarterly U.S. broad inflation and U.S. electricity inflation. 
Here we illustrate that the inversion copulas produce more accurate forecast densities  than the three state space models themselves. In the Supplementary Appendix (Part A) we include a simulation study that shows that even the simplest of our proposed inversion copulas with a flexible margin can greatly increase forecast accuracy. 

\vspace{-10pt}
\subsection{Modeling and Forecasting U.S. Broad Inflation}\label{sec:inflation}
\vspace{-5pt}
We employ our methodology to
model and forecast U.S. inflation 
from 1954:Q1 to 2013:Q4. Inflation is measured by the difference
$y_t=\log(P_t)-\log(P_{t-1})$ in the logarithm of the (seasonally 
adjusted) quarterly GDP price deflator $P_t$,
sourced from the FRED database of the Federal Reserve Bank
of Saint Louis. Figure~\ref{fig:infldata}(a) plots the
time series of the $T=240$ quarterly observations, while
Figure~\ref{fig:margins} plots histograms of the data.
The marginal
distribution of inflation is far from symmetric, with sample skew 1.329 
and kurtosis 4.515, and a
Shapiro \& Wilk~(1965) test for normality is rejected at any meaningful 
significance level. A wide range of time series models have been fitted to
quarterly inflation
data previously (Faust \& Wright~2013; Clark \& Ravazzolo~2015), including the three
state space models considered here. However, these three models--- in fact, 
most time series models
used previously---
have margins that are inconsistent with that observed empirically. We show
that combining each of the three inversion copulas with more flexible margins 
solves this problem, and 
significantly improves the accuracy of the
one-quarter-ahead predictive
densities.

\vspace{-5pt}
\subsubsection{SVUC and InvCop1}\label{sec:inflsv}
\vspace{-5pt}
Stock \& Watson~(2007) suggest using an unobserved
component model with stochastic volatility for U.S. inflation, and it
has become a popular model for this series
(Clark \&~Ravazzolo~2015; Chan~2015). We fit the SVUC model
directly to the inflation data using Bayesian methods.
Table~\ref{tab:psihat}
reports the parameter estimates,
labeled as model `S1'. (Note that this table also reports
the parameter
estimates for all five other models fit to this data.)
The marginal density for inflation implied by the SVUC model
is shown in Figure~\ref{fig:margins}(a)
in red. It
is necessarily symmetric and
inconsistent with that observed empirically.

We therefore employ a copula time series model 
(labeled `C1') 
with copula function InvCop1 and a nonparametric margin,
for which we employ the kernel density estimator (KDE), with the
locally adaptive bandwidth method of Shimazaki \& Shinomoto~(2010).
This copula model allows
for the same serial dependence structure as the SVUC model, 
but with a more accurate
margin.
The estimated margin is 
a smooth asymmetric and
heavy-tailed distribution, and is also
plotted in 
Figure~\ref{fig:margins}(a). Using this for
$g$, 
the copula data $u_t=G(y_t)$ are computed
and plotted in 
Figure~\ref{fig:infldata}(b). This
time series
retains the serial dependence
apparent in the original data.

The copula parameters are estimated using the MCMC scheme,
where the proposals in Step~2 of the sampler
have acceptance rates
between 35\% and 41\%. There is
strong positive correlation in both the level ($\hat \rho_\mu=0.959$) 
and the log-volatilities ($\hat \rho_\zeta=0.789$) of $Z_t$, similar to that
for the SVUC model fit directly to the inflation data.

Figure~\ref{fig:copdensity_comparison}(a) plots the 
marginal 
copula density $c^{(2)}(u_t,u_{t-1}|\hat{\bm{\psi}})$ at the parameter
estimates. There are spikes in the density close to (0,0) and (1,1), 
so that the vertical
axis is truncated at 7 to aid interpretation. The logarithm of the density
is also plotted in panel~(d).
The majority of mass is along the axis between (0,0) to (1,1),
which is due to level
dependence captured by the unobserved
component. However, the
conditional heteroskedasticity also affects the form of the copula, with
mass around points (0,1) and (1,0) and four edges apparent in panel~(d).
Table~\ref{tab:depend}
reports measures of first order serial dependence in $\{Y_t\}$
captured by InvCop1. 
The unobserved mean component results in strong positive overall
dependence, with
$\hat r_1=0.792$. There is high (symmetric) quantile dependence 
$\hat \lambda_1^{++}(0.05)=\hat \lambda^{--}_1(0.05)=0.507$,
consistent with 
the shape of $c^{(2)}$.

\vspace{-5pt}
\subsubsection{MSAR1 \& InvCop2}
\vspace{-5pt}
Amisano \& Fagan~(2013) employ the MSAR1 
model for U.S. inflation,
but only allow $c_1$ and $c_2$ to vary between regimes. 
We extend this study here by fitting 
the more general MSAR1 model (labeled `S2') directly 
to the inflation data
using Bayesian methods.
The implied marginal density
is shown in Figure~\ref{fig:margins}(b) in red, and it is more consistent
with the data than the margin of the SVUC model. 

A time series copula model, with margin $G$
given by the KDE and copula function InvCop2, is also fit 
and labeled as model `C2'. The copula parameters are estimated
using the MCMC scheme, and the proposals
in Step~2 of the sampler have acceptance rates between 13\% and 38\%.
Parameter estimates
for both models S2 and C2 show positive serial dependence and high values
for $p_{11}$ and $p_{22}$. However, the characteristics of the regimes
differ between the two models, and it is shown later that the two models
also
have
very different
predictive distributions

Table~\ref{tab:depend} reports the first order serial dependence metrics
of copula InvCop2. Similar to the other copulas,
there is 
high overall dependence with
$\hat r_1=0.752$, although dependence is highly asymmetric
with $\hat \lambda^{++}_1(0.05)=0.644>\hat \lambda^{--}_1(0.05)=0.272$.
This asymmetry is
visible in
$c^{(2)}$ and $\log(c^{(2)})$, which are plotted in 
Figure~\ref{fig:copdensity_comparison}(b,e).
In contrast, the copulas InvCop1 and InvCop3 do not allow for such
asymmetric quantile dependence.

\vspace{-5pt}
\subsubsection{UCAR4 and InvCop3}
\vspace{-5pt}
As a benchmark,
the UCAR model with $p=4$ is also fitted, and labeled as model `S3'. 
The margin implied by this model is Gaussian, 
and plotted in Figure~\ref{fig:margins}(c) in red. The estimates of
the partial correlations $\bm{\pi}$ suggest that the unobserved component
is Markov order one,
which is consistent with the GDP data being
seasonally-adjusted. To compare,
we also fit a time series copula model (labeled `C3') with 
copula function InvCop3 and $p=4$. To illustrate estimation of a parametric
margin, $G$ is
a skew t distribution (Azzalini \& Capitanio~2003).
The
location ($\xi$),
scale ($\omega$), skew ($\gamma_1$) and kurtosis ($\gamma_2$)
coefficients are used as parameters, so that
$\bm{\theta} =
(\xi,\omega,\gamma_1,\gamma_2)$. 
The joint parameter
posterior is computed using the 
MCMC scheme, where $\bm{\theta}$ is generated in Step~3 of the sampler
using an adaptive random walk proposal.

Table \ref{tab:psihat} reports 
the posterior estimates of both $\bm{\psi}$ and $\bm{\theta}$.
The skew t margin
has high positive skew $\hat \gamma_1=1.565$ and
heavy tails $\hat \gamma_2=7.89$, similar to the KDE.
In contrast to model S3, 
the posterior of $\bm{\pi}$ for model C3 suggests that the unobserved
component is Markov order two. 
Figure~\ref{fig:copdensity_comparison}(c,f)
plots $c^{(2)}$ and $\log(c^{(2)})$ for InvCop3.
This is a bivariate Gaussian copula, and is therefore symmetric
along the axes
(0,1) to (1,0). As with the other copula functions, overall first
order serial dependence
is positive with $\hat r_1=0.789$.
Quantile dependence is symmetric and positive for $\alpha>0$, although
$\lim_{\alpha\rightarrow 0}\lambda_1^{++}(\alpha)=0$ for any Gaussian copula.

\vspace{-5pt}
\subsubsection{Density Forecast Comparison}
\vspace{-5pt}
One-quarter-ahead predictive densities are computed for quarters
$t=2,\ldots,T$ for all six fitted models. Point forecast accuracy 
is measured using the root mean squared error (RMSE). Density forecast accuracy is measured using 
the (negative) logarithm of the predictive score (LP), 
the cumulative rank probability score (CRPS) and the 
tail-weighted CRPS (TW-CRPS). The latter two measures are introduced
in Gneiting \& Raftery~(2007) and Gneiting and Ranjan~(2011), 
and computed directly from the quantile
score as in Smith \& Vahey (2016). The mean values of the metrics are
reported in Table~\ref{tab:mmetric}, where lower values for all metrics indicate
increased accuracy. 
The density forecasts from the copula models C1, C2 and C3 are all more accurate
than those from the corresponding state space models S1, S2 and S3, as measured
by
mean LP, CRPS and TW-CRPS. However,
adopting a copula model results
in less of an improvement in
RMSE.

To show how the predictive distributions differ,
Figure~\ref{fig:pred_moments}(b,d,f)  
plots their standard deviation for
models C1--C3 in blue, and for models S1--S3 in red;
the differences are striking. 
This is
particularly the case for model C3 in panel~(f), where the combination of 
an asymmetric margin with InvCop3
produces heteroskedasticity in the predictive distributions, 
even though the latent state space model is homoskedastic.
A similar feature 
was observed by Smith \& Vahey~(2016) when they fit a Gaussian copula.
The tails of the predictive distributions also differ.
For example, Figure~\ref{fig:deflation} plots the predictive probability of 
deflation for all models. Broad-based deflation is very rare, with only very
mild deflation occurring twice in our data. Yet, 
the state space models can over-estimate
this probability. This is because
the inaccuracy of the left hand tails of the margins of models S1--S3
apparent
in Figure~\ref{fig:margins} also extends to the predictive distributions. 
Figure~\ref{fig:4dens} plots the predictive distributions from models S1 and C1 for four different quarters. It shows 
that the predictive distributions from the copula model do not simply replicate
the asymmetry (or other features) of the marginal
distribution.
Overall,
the best performing model is C1, 
and its inclusion in a
real time forecasting study -- such as those by
Clark \& Ravazzolo~(2015) and Smith \& Vahey (2016) -- is merited.

\vspace{-5pt}
\subsubsection{Computation Times}
\vspace{-5pt}
The MCMC estimation algorithms were implemented in MATLAB using a standard  workstation, 
and computation times varied across the three copula models.
The time to complete 1000 sweeps was
667s, 44s and 94s for models C1, C2 and C3, respectively. The greater computing 
time for  model C1 is because $F_{Z_1}$ requires evaluation of a univariate numerical
integral, as noted in Section~\ref{sec:sv}. In addition, model C3 also involves generation of the 
parameters of the skew t margin. Our empirical results are based on Monte Carlo 
samples of size
20,000, 25,000 and 30,000 iterates for models C1, C2 and C3,
respectively, with a further
5,000, 15,000 and 20,000 iterates discarded for convergence.
While we generated more iterates for the faster schemes, we
found that varying these did not  affect the results
meaningfully.

\vspace{-5pt}
\subsubsection{Maximum Likelihood Estimates}\label{sec:MLE}
\vspace{-5pt}
We report maximum likelihood estimates of the copula parameters for the U.S. broad inflation 
example in Table~\ref{WAtab:psihat} to show that it is possible to estimate the inversion copulas via MLE.
All parameter estimates are obtained using two-stage estimation
(Joe 2005), where the margin $G$ is first estimated and the copula
data $u_t=G(y_t)$ computed for $t=1,\ldots,T$. Conditional upon this
copula data, $\bm{\psi}$ is estimated by maximizing the copula density in
Equation~(2.3). The denominator of this density, as well as the values
of $\bm{z}$, are computed from $f_{Z_1}$, $F_{Z_1}$ and $F_{Z_1}^{-1}$
in the same manner as outlined in the paper.
The numerator $f_Z$ of the copula density can be evaluated for each of our
three inversion copulas by filtering algorithms.
For InvCop1, the likelihood of the
latent SVUC model is computed by 
the bootstrap particle filter (Gordon et al.~1993), in combination with the 
Kalman filter.
For InvCop2, the likelihood of the latent MSAR1 model is evaluated in closed
form using the Hamilton filter for discrete states (see Hamilton 1989).
For InvCop3, the likelihood of the latent state space model is
Gaussian with moments that 
can be computed
using the Kalman filter.
In all cases, $\bm{\psi}$ is constrained as discussed
in Section~3, and maximization employs constrained
optimization as implemented in the Matlab toolbox. The maximum likelihood estimates reported in Table~\ref{WAtab:psihat} are in line with the posteriors in Table~\ref{tab:psihat}.

\vspace{-10pt}
\subsection{Modeling and Forecasting U.S. Electricity Inflation}
\vspace{-5pt}
To confirm that our methodology applies to other data, we consider inflation in U.S. 
electricity prices between 1952:Q1 and 2015:Q4 as a second empirical example. 
The data are differences in the logarithm of the (seasonally-adjusted) quarterly electricity consumer price index of all urban consumers.
This series is
produced by the U.S. Bureau of Labor Statistics, and available from the FRED database.
The time series plots of
the $T=256$ observations of the data and the associated copula data are given in Figure 1 of the Supplementary Appendix. The marginal distribution of the data is highly non-Gaussian,
and we fit the same six models to this item-specific 
inflation data, as we do the broad
inflation measure in Section~\ref{sec:inflation}. 
Even though the data differ, the three state space models S1, S2 and S3
remain attractive time series models for item-specific inflation.

Figure~\ref{WAfig:margins} plots the histogram of the data in every panel.
As with the broader inflation measure, electricity inflation is positively
skewed. It also exhibits an excentuated peak around 0\%. 
The marginal distributions of the fitted state space models S1, S2 and S3 are plotted
in red in panels~(a,b,c), respectively. The symmetric margins of S1 (the
SVUC model) and S3 (the UCAR4 model) are highly inconsistent with the data.
Also plotted in blue in panels~(a,b) 
is the 
adaptive KDE estimate --- which is far from symmetric --- as is
the margin of both copula models C1 and C2. Panel~(c) plots the posterior estimate
of the skew t distribution, which is the margin of copula model C3. It is positively skewed
and heavy-tailed. In each case, the margins of the copula models are 
more accurate than
those implied by the respective state space models.

The copula densities from all three models are similar to those in the broad inflation case, with the bivariate marginal copulas  $c^{(2)}(u_t,u_{t-1})$ showing strong positive dependence (see Figure~2 of the Supplementary Appendix).
This is unsurprising because electricity consumption
is a major component of economic output. 
As in Section~\ref{sec:inflation}, we compute the one-quarter-ahead predictive distributions
for all six models at times $t=2,\ldots,T$. 
Table~\ref{WAtab:metrics}
reports the accuracy metrics, and in each case the copula time series
models out-perform
their equivalent state space models using every metric. 
One-sided t-tests of the CRPS and log-score
suggest that these differences are statistically significant.
As with the analysis of broad inflation,
the flexible modelling of the highly asymmetric
margin increases the quality of the fitted time series model and 
the accuracy of these predictive densities. Illustration of the stark differences between the one-quarter-ahead predictive densities from
the state space models, and their equivalent
copula models, is given in Figure~3 of the Supplementary Appendix. 

%% file: sect5.tex
\vspace{-12pt}
\section{Discussion}\label{sec:disc}
\vspace{-5pt}
This paper proposes a new class of copulas for capturing
serial dependence. 
They are constructed from inversion of
a general nonlinear state space model, so that the potential range of
dependence structures that they can produce
is incredibly broad.
A major insight is that such copulas
can be very different than those that are widely used for capturing
cross-sectional dependence. The latter include elliptical 
and vine copulas, which have also been used previously to capture serial dependence;
see 
Joe~(1997), Beare~(2010),
Smith et al.~(2010) and Loaiza-Maya et al.~(2017) 
for examples. Yet, the three inversion copulas
studied in detail highlight the wider serial dependence structures 
that can be captured by our approach. 

As with the likelihood of the
underlying state space model, in general
the density of the inversion copula
cannot be expressed in closed form.
However,
an important insight is that
existing methods for evaluating the likelihood of the latent
state space model can also be employed in the copula context.
While we employ MCMC samplers to evaluate the posterior,
other existing
methods
can also be used
to compute the numerator of the copula density, as we do  
in Section~\ref{sec:MLE}. Either way, a major
computational challenge 
is the repeated evaluation of
the quantile 
$F_{Z_t}^{-1}$ and
density $f_{Z_t}$ functions at the $T$ observations.
When the latent state space model is stationary, 
we show how this can be achieved using
spline approximations to $F_{Z_1}^{-1}$ and $\log(f_{Z_1})$ outlined
in Appendix~A. These approximations
are highly
accurate in our examples, fast to derive,
and can be employed with even very large values of $T$ in practice.

Recently, Oh \& Patton~(2015) construct an inversion copula
from a flexible parametric distribution formed through marginalization
over a low-dimensional vector of latent factors. 
This latent factor model can be written
in state space form, where the factors are a static state
vector $\bm{X}_t$ that does not vary with $t$ at
Equation~(\ref{eq:trans}). Full likelihood-based
estimation can then be undertaken
using the Bayesian MCMC methods discussed here, 
providing a better alternative to the moment-based method
suggested by Oh \& Patton~(2015).
While these factor copulas are unsuitable for serial dependence,
Oh \& Patton~(2015) show they can capture high-dimensional cross-sectional
dependence well.

An interesting result is that the inversion copula of 
a Gaussian linear state space model is a Gaussian copula. 
Therefore, in this special 
case, the likelihood is available in closed form. Nevertheless,
estimation using
simulation 
methods can still prove efficient, just as it is
for the latent state space model itself. While all three of our
example inversion copulas are Markov and stationary, copulas
can also be derived from non-stationary latent state space models. 
The resulting 
time series copula model is also non-stationary, but with a time
invariant margin. Such copula
models are an interesting topic for further study, although a new 
approach to computing $F_{Z_t}^{-1}$ and $f_{Z_t}$ efficiently is needed. 
A second interesting extension is to employ the proposed
time series inversion copulas to capture serial dependence in discrete 
data. Here, the copula remains unchanged, but $G$ would be
a discrete distribution function.
The model can be estimated using Bayesian 
data augmentation, as discussed in Pitt, Chan \& Kohn~(2006)
and Smith, Gan \& Kohn~(2012).
This would require
$\bm{z}$ to be generated as an additional step in the sampling scheme
in Section~\ref{sec:copmod}.

A third highly useful extension is
to construct the inversion copula of a multivariate state space 
model. If the dimension of the multivariate time series is $m$, then 
the resulting $Tm$-dimensional copula
captures both cross-sectional and serial dependence
jointly. This would provide an alternative to the vine copula models of 
Smith~(2015), Beare \& Seo~(2015) and Loaiza-Maya et al.~(2017) for this case.
While the extension is straightforward in principle, 
implementation relies on the ability to evaluate the
univariate marginal distribution functions (and their inverses) for each series
of the latent state space model.

To show our methodology can improve the quality of forecast densities, we use it to model and forecast quarterly U.S. broad inflation and U.S. electricity inflation, which is
an important problem in empirical macroeconomics (Faust \& Wright~2013). 
We employ inversion copulas constructed
from three state space models used previously
for this series.
When combined with highly asymmetric and heavy-tailed nonparametric or 
flexible marginal distributions, the predictive distributions from
the resulting copula time series models are more accurate than those of
the state space models themselves. This is because these state space
models have rigid margins, which are very far from that observed for 
inflation empirically--- a problem resolved by the copula models.

%% file: acknowl.tex
\noindent{\large {\bf Acknowledgments}}\\
The work of Michael Smith was supported by
Australian Research
Council Grant FT110100729. 

%% file: append.tex
\appendix
\noindent {\bf \Large{Appendix}}\\
\vspace{-20pt}

\noindent {\large {\bf Part~A: Fast Evaluation of the Marginal Density and Quantile Function}}\\
\noindent 
This part of the appendix outlines how to efficiently evaluate
the quantile function $F_{Z_1}^{-1}$
and
the logarithm
of the marginal
density $\log(f_{Z_1}(z))$ for a stationary nonlinear state space model.
For both, we
use spline interpolations
based on their values at $N$ absciassae, where
we set $N=100$ in practise.
The advantage of such spline-based approximations is that they are
highly accurate (between 5 and 9 decimal places in our empirical
work), 
yet are fast to compute at the $T$ observations once
the interpolation is complete-- even for large values of $T$.

We use a uniform grid for the $N$
quantile
function values
$\{q_1,\ldots,q_N\}$, which have correpsonding probability values
$\{p_1,\ldots,p_N\}$. 
We set $p_1=0.0001$ and $p_N=0.9999$, so that
the function is approximated far into the tails of the distribution.
The following steps obtain the 
points at which the interpolations are made.
\begin{itemize}
\setlength\itemsep{-0.5em}
\item[1.] Set $p_1=0.0001$ and $p_N=0.9999$, and evaluate both
$q_1=F_{Z_1}^{-1}(p_1)$ and $q_N=F_{Z_1}^{-1}(p_N)$
using a root finding algorithm.
\item[2.] Set step size to $\delta=(q_N-q_1)/(N-1)$, and a construct
uniform grid as $q_i=q_1+(i-1)\delta$, for $i=2,\ldots,N$.
\item[3.] For $i=1,\ldots,N$ (in parallel): \\
\indent \indent 3a. Compute $p_i=F_{Z_1}(q_i)$\\
\indent \indent 3b. Compute $b_i=\log(f_{Z_1}(q_i))$
\end{itemize}

We then interpolate the points $\{(p_i,q_i);i=1,\ldots,N\}$
and $\{(q_i,b_i);i=1,\ldots,N\}$ using splines. We employ
natural cubic smoothing splines 
using the (fast and efficient)
spline toolbox in MATLAB, although other fast interpolating methods
could also be employed. 
Notice that
numerical
inversion of $F_{Z_1}$ is undertaken above only twice in Step~1. Moreover,
$F_{Z_1}$ and $f_{Z_1}$ are evaluated
only $N$ times in step~3, something
that can also be undertaken parallel. 

Once the coefficients of the splines are obtained, the log-density and quantile function can be evaluated quickly
at even very large number of points $\{z_1,\ldots,z_n\}$.
If $F_{Z_1}$ is also symmetric (as in Section~\ref{sec:sv}), then
$F_{Z_1}^{-1}(1-u)=-F_{Z_1}^{-1}(u)$ for $0\leq u\leq 1/2$, 
$f_{Z_1}(z)=f_{Z_1}(-z)$ and
$F_{Z_1}(-z)=1-F_{Z_1}(z)$ for $z\geq 0$. These identities can be exploited
to reduce the number of computations at Steps~1 and~3 by one half, further
speeding the algorithm.
 
To illustrate the effectiveness of the method, we consider the approximations
to $F_{Z_1}^{-1}$ and $\log(f_{Z_1})$ 
for InvCop1 when $\bm{\psi}$ equals the
posterior mean in the inflation study
in Section~\ref{sec:inflsv}.
Plots of the approximations (see Figure~4 of the supplementary material) are visually
indistinguishable from the true functions,
which can be evaluated (slowly) using numerical methods. The integrated
absolute difference between the approximate and true functions are 
$1.282\times 10^{-6}$ and $2.253\times 10^{-10}$ for the quantile and log-density,
respectively, so that the approximations are 
very accurate.
Computation of both approximations, and their evaluation at the $T=240$ observations, 
takes only 0.29s using MATLAB on a standard four core desktop.

\noindent {\bf \large{Part~B: Bivariate Marginal Copulas}}\\
\noindent 
This part of the appendix shows how to evaluate the bivariate marginal
copula density
\[
c^{(2)}(u_1,u_2;\bm{\psi})=
\frac{f_Z^{(2)}(z_1,z_2|\bm{\psi})}{f_{Z_1}(z_2|\bm{\psi})f_{Z_1}(z_1|\bm{\psi})}\,,
\]
for InvCop1 and InvCop2. In both cases,
the univariate marginal density
$f_1(z|\bm{\psi})$ can be computed readily as in Part~A above.
Computation of $f_Z^{(2)}$ is outlined
separately for each case below. 

\noindent {\bf For InvCop1}\\
\noindent
For the SVUC model with the parameter constraints, the bivariate
density 
\[f^{(2)}_Z(z_1,z_2|\bm{\psi})=\int\int\int\int 
	\phi_2\left(\bm{z};\bm{\mu},V(\bm{\zeta})\right)\,
\phi_2(\bm{\mu};\bm{0},S_\mu)\mbox{d}
\bm{\mu}\,\phi_2(\bm{\zeta};(\bar \zeta ,\bar \zeta),S_\zeta) \mbox{d}\bm{\zeta}\,,
\]
where
$\bm{z}=(z_1,z_2)$, $\bm{\mu}=(\mu_1,\mu_2)$, $\bm{\zeta}=(\zeta_1,\zeta_2)$,
\[
V(\bm{\zeta})=\left[\begin{array}{cc} \exp(\zeta_1)& 0\\ 0& \exp(\zeta_2) \end{array}\right],\,
S_\zeta=s_\zeta^2\left[\begin{array}{cc} 1& \rho_\zeta\\ \rho_\zeta& 1 \end{array}\right],\,
S_\mu=s_\mu^2\left[\begin{array}{cc} 1& \rho_\mu\\ \rho_\mu& 1 \end{array}\right],\, 
\]
and $\phi_2(\bm{x};\bm{a},\Omega)$ is a bivariate normal density
with mean $\bm{a}$ and variance $\Omega$
evaluated at point $\bm{x}$.
The inner two integrals in $\bm{\mu}$ of this 4-dimensional integral
can be computed analytically by recognising a bivariate normal.
Then, by recognising a second bivariate normal in
$\bm{z}$, the density can be written as:
\[f(z_1,z_2|\bm{\psi})=\int\int \phi_2(\bm{z};\bm{0},W(\bm{\zeta}))\phi_2(\bm{\zeta};(\bar \zeta,\bar \zeta),S_\zeta)
\mbox{d}\bm{\zeta}\,,
\]
where $W(\bm{\zeta})=(S_\mu+V(\bm{\zeta}))$. This bivariate integral can be
computed numerically.

\noindent {\bf For InvCop2}\\
\noindent 
For the MSAR1 model with the parameter constraints, the bivariate density 
is the mixture of four bivariate Gaussians
\begin{eqnarray*}
f^{(2)}_Z(z_1,z_2|\bm{\psi})=\sum_{i=1,2}\sum_{j=1,2}\phi_2 
\left( (z_1,z_2);\bm{\mu}_{ij},
S_{ij} \right)p_{ij}\pi_i\,,
\end{eqnarray*}
where $\bm{\mu}_{ij}=\left(\mu_i,\mu_j\right)$ and
\[
S_{ij}=\begin{bmatrix} s^2_i & \rho_j s^2_i \\ \rho_j s^2_i & s^2_j \end{bmatrix}\,.
\] 

%% file: refs.tex
\vspace{-20pt}
\section*{References}
\vspace{-10pt}
\parindent=0pt              
\begin{trivlist}
\item[]
Almeida, C. and C. Czado, (2012). 
`Bayesian inference for stochastic time-varying copula models',
{\em Computational Statistics \& Data Analysis}, 56, 1511--1527.
\item[]
Amisano, G. \& G. Fagan, (2013). `Money growth and inflation: 
A regime switching approach', {\em Journal of International Money
and Finance}, 33, 118--145.
\item[]
Azzalini, A. \& A. Capitanio,~(2003). `Distribution generated by perturbation
of symmetry with emphasis on a multivariate skew t distribution',
{\em Journal of the Royal Statistical Society}, Series B, 65, 367--389.
\item[]
Barndorff-Nielsen, O. \& G. Schou, (1973). `On the Parameterization of Autoregressive
Models by Partial Autocorrelations', {\em Journal of Multivariate Analysis}, 3, 408--419.
\item[]
Beare, B. K., (2010). `Copulas and temporal dependence', {\em Econometrica},
78, 395--410.
\item[]
Beare, B. K., \& J. Seo~(2015). `Vine copula specifications for stationary 
multivariate Markov chains', {\em Journal of Time Series Analysis}, 
36, 228--246.
\item[]
Brockwell, P.J. \& R.A. Davis,~(1991), {\em Time Series: Theory and Methods},
2nd Ed., NY: Springer.
\item[]
Chan, J.C.C (2015). `The Stochastic Volatility in Mean Model with
Time-Varying Parameters: An Application to Inflation Modeling',
{\em Journal of Business and Economic Statistics}, to appear.
\item[]
Chan, J.C.C. \& C.Y.L. Hsiao, (2014). 
`Estimation of Stochastic Volatility Models with Heavy Tails and Serial 
Dependence', in {\em Bayesian Inference in the Social Sciences},
Eds. I. Jeliazkov and X.-S. Yang, Wiley: NJ.
\item[]
Clark, T. E. \& F. Ravazzolo, (2015). 
`Macroeconomic Forecasting Performance under Alternative Specifications of 
Time-varying Volatility', {\em Journal of Applied Econometrics}, 30, 
551--575.
\item[]
Creal, D.D. \& R.S. Tsay, (2015).
`High dimensional dynamic stochastic copula models', {\em Journal of Econometrics},
189, 335--345.
\item[]
DeJong, D.N., R. Liesenfeld, G.V. Moura, J.-F. Richard, H. Dharmarajan, (2013).
`Efficient Likelihood Evaluation of State-Space Representations',
{\em Review of Economic Studies}, 30, 538--567. 
\item[]
De Lira Salvatierra, I. \& A.J. Patton~(2015), `Dynamic copula models and high
frequency data', {\em Journal of Empirical Finance}, 30, 120--135.
\item[]
Demarta, S. \& A. McNeil, (2005).
`The t-copula and related copulas', {\em International
Statistical Review}, 73, 111--129.
\item[]
Durbin, J. \& S.J. Koopman, (2012). {\em Time Series Analysis by
State Space Methods}, 2nd Ed., OUP.
\item[]
Embrechts, P., A. McNeil \& D. Straumann, (2001).
`Correlation and dependency in
risk management: properties and pitfalls', in 
M. Dempster \& H. Moffatt, (Eds.)
{\em Risk Management: Value at Risk and Beyond}
Cambridge University Press, 176--223.
\item[]
Faust, J. \& J. Wright, (2013). `Forecasting Inflation', in G. Elliot \&
A. Timmermann (eds.), {\em Handbooks in Economics}, Vol. 2A., 3--56. 
\item[]
Frees, E.W. \& P. Wang, (2006). `Copula credibility for aggregate loss
models', {\em Insurance: Mathematics and Economics}, 38, 360--373.
\item[]
Hafner, C.M. \& H. Manner, (2012). `Dynamic stochastic copula models:
estimation, inference and applications', {\em Journal of Applied Econometrics},
27, 269--295.
\item[]
Hamilton, J.D., (1994). {\em Time Series Analysis}, Princeton University
Press: NJ.
\item[]
Gneiting, T. \& A.E. Raftery,~(2007). `Strictly Proper Scoring Rules,
Prediction and Estimation', {\em Journal of the American Statistical
Association}, 102, 359--378.
\item[]
Gneiting, T. \& R. Ranjan,~(2011). `Comparing Density Forecasts using Threshold- and Quantile-Weighted Scoring Rules', {\em Journal of Business and Economic Statistics}, 29, 411--422.
\item[]
Godsill, S., A. Doucet \& M. West~(2004). `Monte Carlo Smoothing for 
Nonlinear Time Series', {\em Journal of the American Statsitical Association},
99, 156--168.
\item[]
Gordon, N.J., D.J. Salmond and A.F.M. Smith (1993). `A Novel Approach to Non-Linear and Non-Gaussian Bayesian State Estimation',
{\em IEEE Proceedings, F140}, 107--133.
\item[]
Hamilton, J.D. (1989). `A New Approach to the Economic Analysis of Nonstationary Time Series and the Business Cycle',
{\em Econometrica}, 57, 357--384.
\item[]
Joe, H., (1997). {\em Multivariate Models and Dependence Concepts}, Chapman
and Hall.
\item[]
Joe, H. (2005). `Asymptotic Efficiency of the Two-Stage Esimation Method for Cupula-Based Models',
{\em Journal of Multivariate Analysis}, 94, 401--419.
\item[]
Jungbacker, B. \& S.J. Koopman~(2007).
`Monte Carlo Estimation for Nonlinear Non-Gaussian State Space Models',
{\em Biometrika}, 94: 4, 827-839.
\item[]
Kauermann, G., C. Schellhase, D. Ruppert~(2013). `Flexible Copula
Density Estimation with Penalized Hierarchical B-splines',
{\em Scandinavian Journal of Statistics: Theory and Applications},
40, 685--705.
\item[]
Kim, S., N. Shephard \& S. Chib, (1998). `Stochastic Volatility: Likelihood
Inference and Comparison with ARCH Models', {\em Review of Economic Studies},
65, 361--393.
\item[]
Lambert, P. \& F. Vandenhende, (2002). `A copula-based model for
multivariate non-normal longitudinal data: analysis of a dose titration safety
study on a new antidepressant', {\em Statistics in Medicine}, 21, 3197--3217.
\item[]
Ljung, L. (1999). {\em System Identification: Theory for the User},
2nd Ed., Prentice-Hall.
\item[]
Loaiza-Maya, R., M.S. Smith \& W. Maneesoonthorn~(2017). `Time Series Copulas
for Heteroskedastic Data', {\em Journal of Applied Econometrics}, forthcoming.
\item[]
L\"{u}tkepohl, H., (2006). {\em New Introduction to Multiple Time Series
Analysis}, Springer-Verlag: Berlin.
\item[]
Oh, D.H. \& A.J. Patton~(2015). `Modelling Dependence in High Dimensions 
with Factor Copulas', {\em Journal of Business and Economic Statistics}, forthcoming.
\item[]
Nelsen, R., (2006), {\em An Introduction to Copulas}. 2nd ed., New York, Springer.
\item[]
Patton, A.J., (2006). `Modelling Asymmetric Exchange Rate Dependence', {\em International Economic Review}, 47, 527--556. 
\item[]
Patton, A.J., (2012). `A Review of Copula Models for Economic Time
Series', {\em Journal of Multivariate Analysis}, 110, 4--18.
\item[]
Pitt, M., D. Chan \& R. Kohn~(2006). 
`Efficient Bayesian Inference for Gaussian Copula Regression Models', 
{\em Biometrika}, 93, 537--554.
\item[]
Richard, J.F. \& W. Zhang, (2007). `Efficient high-dimensional importance sampling',
{\em Journal of Econometrics},
141, 1385--1411.
\item[]
Scharth, M. \& Kohn, R.,~(2016). `Particle Efficient Importance Sampling',
{\em Journal of Econometrics}, 190, 133--147.
\item[]
Shapiro, S. \& M. Wilk, (1965). `An Analysis of Variate Test for Normality
(Complete Samples)', {\em Biometrika}, 52, 3-4, 591--611.
\item[]
Shephard, N. \& M. Pitt,~(1997). `Likelihood analysis of non-Gaussian 
measurement time series', {\em Biometrika}, 84:3, 653--667.
\item[]
Shih, J.H. \& T.A. Louis, (1995). `Inferences on the association parameter in copula models
for bivariate survival data', {\em Biometrics}, 51, 4, 1384--99.
\item[]
Shimazaki, H. \& S. Shinomoto, (2010). `Kernel bandwidth optimization in
spike rate estimation', {\em Journal of Computational Neuroscience}, 
29, 171--182.
\item[]
Smith, M.S., (2015). `Copula modelling of dependence in multivariate time
series', {\em International Journal of Forecasting}, 31, 815--833.
\item[]
Smith, M.S., Q. Gan \& R. Kohn~(2012). `Modelling Dependence using Skew t
Copulas: Bayesian Inference and Applications', {\em Journal of Applied
Econometrics}, 27, 500--522.
\item[]
Smith, M., A. Min, C. Almeida \& C. Czado, (2010). 
`Modeling Longitudinal Data using a Pair-Copula Decomposition 
of Serial Dependence',
{\em Journal of  the American Statistical Association}, 
105, 492, 1467-1479.
\item[]
Smith, M.S. \& S.P. Vahey, (2016). 
`Asymmetric density forecasting of U.S. macroeconomic variables
using a Gaussian copula model of cross-sectional and serial
dependence', {\em Journal of Business and Economic Statistics}, 34:3, 416--434. 
\item[]
Song, P. (2000).
`Multivariate Dispersion Models Generated from Gaussian
Copula', {\em Scandinavian Journal of Statistics}, 27, 305--320.
\item[]
Stock, J.H. and M.W. Watson,~(2007).
`Why Has U.S. Inflation Become Harder to Forecast?',
{\em Journal of Money, Credit and Banking}, 39, 3--33.
\item[]
Stroud, J.R., P. M\"uller \& N.G. Polson,~(2003).
`Nonlinear State-Space Models with State-Dependence Variances',
{\em Journal of the American Statistical Association}, 98:462, 377--386.
\item[]
Tsukahara, H. (2005). `Semiparametric estimation in copula models',
{\em The Canadian Journal of Statistics}, 33, 357--375.
\end{trivlist}

%% file: tabs.tex
\begin{table}[th]
\begin{center}
\begin{tabular}{lccccc}\hline
& &$U_t=G(Y_t)$ & &$Z_t=F_{Z_t}^{-1}(U_t)$ & \\
Process &$\{Y_t\}_{t=1}^T$ &$\longrightarrow$ &$\{U_t\}_{t=1}^T$ &$\longrightarrow$ &$\{Z_t\}_{t=1}^T$ \\
	Domain &${\cal S}_Y\subset {\cal R}^T$ &$\longrightarrow$ &$[0,1]^T$ &$\longrightarrow$ &${\cal S}_Z\subset {\cal R}^T$ \\
	Joint CDF &$F_Y(\bm{y})$ &$\longrightarrow$ &$C(\bm{u})$ &$\longrightarrow$ &$F_Z(\bm{z})$ \\
Marginal CDFs &$G(y_t)$ &$\longrightarrow$ &Uniform &$\longrightarrow$ &$F_{Z_t}(z_t)$ \\ \hline
\end{tabular}
\end{center}
\caption{Depiction of the transformations underlying an inversion
copula model when $Y_t$ is continuous-valued.}
\label{tab:icd}
\end{table}

\begin{table}[h]
\begin{center}
{\small	
\begin{tabular}{ccccccc} \hline \hline  \noalign{\smallskip}
\multicolumn{6}{l}{{\em Copula Time Series Models}} \\ \hline \noalign{\smallskip}
\multicolumn{3}{l}{Model C1: InvCop1 \& KDE Margin}  & & & \\  \cline{1-3} \noalign{\smallskip}
$\rho_\mu$ &$\sigma^2_\mu$ & $\rho_\zeta$ &$\sigma^2_\zeta$ &$\bar{\zeta}$ &$\bar{\mu}$ \\ 
0.959 &0.066 &0.789 &0.603 &-2.573 &0\\
{\footnotesize (0.94,0.97)} &{\footnotesize (0.04, 0.10)} &{\footnotesize (0.48,0.96)} &{\footnotesize (0.08,1.69)} &{\footnotesize (-3.26,-1.94)} & -- \\  \noalign{\smallskip} 
\multicolumn{3}{l}{Model C2: InvCop2 \& KDE Margin} & & & \\  \cline{1-3} \noalign{\smallskip}
$c_1$ &$\rho_1$ &$\sigma^2_1$ &${s^2_1}$ &$p_{11}$ &$\pi_1$ &\\ 
$-0.554$ &0.207 &0.752 &0.815 &0.865 &0.354\\
{\footnotesize (-0.85,-0.33)}&{\footnotesize (-0.11,0.46)} &{\footnotesize (0.53,0.98)} &{\footnotesize (0.57,1.06)} &{\footnotesize (0.75,0.95)} &{\footnotesize (0.20,0.46)} &\\ 
$c_2$ &$\rho_2$ &$\sigma^2_2$ &${s^2_2}$ &$p_{22}$ &$\pi_2$ &\\  
0.040 &0.914 &0.199 &1.239 &0.930 &0.646&\\
{\footnotesize (0.01,0.08)} &{\footnotesize (0.88,0.94)} &{\footnotesize (0.16,0.25)} &{\footnotesize (0.98,1.53)} &{\footnotesize (0.88,0.97)} &{\footnotesize (0.54,0.80)}& \\  \noalign{\smallskip}
\multicolumn{3}{l}{Model C3: InvCop3 \& Skew t Margin} & & & \\  \cline{1-3} \noalign{\smallskip}
$\pi_1$ &$\pi_2$ & $\pi_3$ &$\pi_4$ &$\sigma^2_\mu$  &$\sigma^2$\\ 
0.866 & 0.371 & -0.037 & 0.113 & 0.181 & 0.088 \\
{\footnotesize (0.81,0.93)} &{\footnotesize (0.16,0.55)} &{\footnotesize (-0.41,0.22)} &{\footnotesize (-0.14,0.31)} &{\footnotesize (0.08,0.28)}&{\footnotesize (0.01,0.16)}\\ 
 $\bar{\mu}$ &$\xi$ & $\omega$ & $\gamma_1$ & $\gamma_2$ & \\
0 & 0.202 & 0.549 & 1.565 & 7.890 & \\
-- &{\footnotesize (0.16,0.25)} & {\footnotesize (0.45,0.66)} & {\footnotesize (1.54,1.59)} & {\footnotesize (7.65,8.14)} & \\ \hline \noalign{\smallskip} 
\multicolumn{6}{l}{{\em State Space Models}} \\ \hline   \noalign{\smallskip}
\multicolumn{3}{l}{Model S1: SVUC} & & & \\  \cline{1-3} \noalign{\smallskip}
$\rho_\mu$ &$\sigma^2_\mu$ & $\rho_\zeta$ &$\sigma^2_\zeta$ &$\bar{\zeta}$ &$\bar{\mu}$ \\ 
 0.976 &0.014 &0.904 &0.320 &-3.896 &0.645 \\
{\footnotesize (0.95,0.99)} &{\footnotesize (0.01,0.02)} &{\footnotesize (0.77,0.98)} &{\footnotesize (0.07,0.82)} &{\footnotesize (-4.83,-3.02)} &{\footnotesize (-0.03,1.22)} \\   \noalign{\smallskip} 
\multicolumn{3}{l}{Model S2: MSAR1} & & & \\  \cline{1-3} \noalign{\smallskip}
$c_1$ &$\rho_1$ &$\sigma^2_1$ &${s^2_1}$ &$p_{11}$ &$\pi_1$ &\\  
0.438 &0.687 &0.124 &0.253 &0.960  &0.367 &\\
{\footnotesize (0.19,0.72)} &{\footnotesize (0.53, 0.83)} &{\footnotesize (0.09,0.16)} &{\footnotesize (0.17,0.39)} &{\footnotesize (0.91,0.99)} &{\footnotesize (0.18,0.48)}\\
$c_2$ &$\rho_2$ &$\sigma^2_2$ &${s^2_2}$ &$p_{22}$ &$\pi_2$ &\\ 
0.247 &0.479 &0.039 &0.076 &0.979 &0.633 &\\
{\footnotesize (0.06,0.35)} &{\footnotesize (0.19, 0.89)} &{\footnotesize (0.02,0.06)} &{\footnotesize (0.02,0.23)} &{\footnotesize (0.96,0.99)} &{\footnotesize (0.52,0.82)} \\  \noalign{\smallskip} 
\multicolumn{3}{l}{Model S3: UCAR4} & & & \\  \cline{1-3} \noalign{\smallskip}
$\pi_1$ &$\pi_2$ & $\pi_3$ &$\pi_4$ &$\sigma^2_\mu$  &$\sigma^2$ \\ 
0.920 &0.333 &-0.152 &0.040 &0.037 &0.023 \\
{\footnotesize (0.82,0.98)} &{\footnotesize (-0.03,0.63)} &{\footnotesize (-0.63,0.25)} &{\footnotesize (-0.35,0.30)} &{\footnotesize (0.01,0.07)} &{\footnotesize (0.00,0.04)} \\ 
 $\bar{\mu}$& & & & & \\ 
 0.859& & & & & \\
 {\footnotesize (0.337,1.387)}& & & & \\ \noalign{\smallskip} 
\hline\hline
\end{tabular}
}
\end{center}
\caption{Posterior parameter estimates for the six models fit to the U.S. broad inflation data.
The top half of the table reports estimates for the three inversion copulas
constructed from latent state space models. 
The bottom half reports estimates for 
the same state space models fit directly to the data.
The posterior mean of each parameter is reported,
along with 90\%
posterior probability intervals below.
For InvCop3, the parameters of the jointly estimated 
skew t margin are also reported.
The identification constraint
$\bar\mu=0$ occurs in both InvCop1 and InvCop3.}
\label{tab:psihat}
\end{table}

\begin{table}[h]
\begin{center}
\begin{tabular}{lcccccc} \hline \hline
&\multicolumn{6}{l}{Dependence Metric} \\ \cline{2-7} \noalign{\smallskip}
&$r_1$ &$\tau_1$ &$\lambda_1^{\tiny --}(0.01)$ &$\lambda_1^{\tiny --}(0.05)$ &$\lambda_1^{\tiny ++}(0.01)$ 
 &$\lambda_1^{\tiny ++}(0.05)$ \\ \cline{2-7}  \noalign{\smallskip}
{\em Copula}  & & & & \\
InvCop1 &0.792 &0.611 &0.335 &0.507 &Sym &Sym \\ 
InvCop2 & 0.752 &0.577 &0.191 &0.272 &0.554 &0.644  \\ 
InvCop3 &0.789 &0.578 &0.362 &0.482 &Sym &Sym  \\ 
\hline\hline
\end{tabular}
\end{center}
\caption{Posterior means of first order serial dependence measures for each of the three 
inversion copulas fit to the U.S. broad inflation data.
The copulas InvCop1 and InvCop3
have symmetric tail dependence, while
InvCop2 has asymmetric tail dependence.}
\label{tab:depend}
\end{table}

\begin{table}[h]
\begin{center}
\begin{tabular}{lcccc} \hline \hline
Model & LP & CRPS &TW-CRPS &RMSE \\ \hline
{\em Copula Time Series Models}  & &  & &\\
C1: KDE Margin \& InvCop1 &$-0.0266^\star$ &$0.1393^{\star \star}$ &$0.0308^\star$ &$0.2641^{\star}$ \\
C2: KDE Margin \& InvCop2 &$-0.0086^{\star \star}$ &0.1440 &0.0322 &0.2746 \\
	C3: Skew t Margin \& InvCop3 &0.0409 &$0.1412^{\star}$ &$0.0312^{\star\star}$ &0.2638\\ \hline
{\em State Space Models} & & & &\\
S1: SVUC &0.0166 &0.1439 &0.0318 &0.2712\\
S2: MSAR1 &0.0424 &0.1485 &0.0330 &0.2827 \\
S3: UCAR4 &0.0896 &0.1445 &0.0328 &0.2618\\  \hline \hline
\end{tabular}
\end{center}
\caption{Summary of the accuracy of the one-step-ahead predictive
distributions
for the six models fit to the 
U.S. broad inflation data. The metrics are the mean (negative) logarithm predictive score (LP),
the mean cumulative rank probability score (CRPS), the mean tail-weighted
CRPS (TW-CRPS), and the 
root mean squared error (RMSE). Lower values of all metrics indicate improved
accuracy.
The first three models employ inversion copulas
constructed from latent state space models, along with an asymmetric margin. 
The bottom three
are the same state space models fit directly to the data. When the result
for a copula model is statistically significantly better than the corresponding
state space model, it is indicated with `$\star$' at 10\% significance level or `$\star\star$' at 5\% significance level.}
\label{tab:mmetric}
\end{table}

\begin{table}[htbp]
	\begin{center}
		\begin{tabular}{ccccccc} \hline \hline  \noalign{\smallskip}
			\multicolumn{6}{l}{\em Model C1: KDE Margin \& InvCop1}  \\  \cline{1-6} \noalign{\smallskip}
			$\rho_\mu$ &$\sigma^2_\mu$ & $\rho_\zeta$ &$\sigma^2_\zeta$ &$\bar{\zeta}$ &$\bar{\mu}$ \\ 
			0.948 &0.062 &0.696 &0.567 &$-1.495$ &0\\
			\multicolumn{6}{l}{\em Model C2: KDE Margin \& InvCop2}  \\  \cline{1-6} \noalign{\smallskip}
			$c_1$ &$\rho_1$ &$\sigma^2_1$ &${s^2_1}$ &$p_{11}$ &$\pi_1$ &\\  
			$-0.178$ &0.338 &0.750 &0.846 &0.832 &0.145\\
			$c_2$ &$\rho_2$ &$\sigma^2_2$ &${s^2_2}$ &$p_{22}$ &$\pi_2$ &\\  
			0.004 &0.919 &0.159 &1.025 &0.972 &0.855\\
			\multicolumn{6}{l}{\em Model C3: Skew t Margin \& InvCop3} \\  \cline{1-6} \noalign{\smallskip}
			$\pi_1$ &$\pi_2$ & $\pi_3$ &$\pi_4$ &$\sigma^2_\mu$  &$\sigma^2$\\ 
			0.954 &0.393 &$-0.542$ &0.276 &0.04 &0.144\\
			$\bar{\mu}$ &$\xi$ & $\omega$ & $\gamma_1$ & $\gamma_2$ & \\
			0 &0.172  &0.569 &1.566 &7.891 & &\\ \hline\hline
		\end{tabular}
	\end{center}
	\caption{
		Maximum likelihood parameter estimates for the three inversion copulas
		constructed from latent state space models and fit to the U.S. broad inflation data.
		For InvCop3, the parameters of the estimated 
		skew t margin are also reported.
		The identification constraint
		$\bar\mu=0$ occurs in both InvCop1 and InvCop3.}
	\label{WAtab:psihat}
\end{table}

\begin{table}[htbp]
	\begin{center}
		\begin{tabular}{lccc} \hline \hline
			Model & LP & CRPS & RMSE \\ \hline
			{\em Copula Time Series Models}  & & &\\
			C1: KDE Margin \& InvCop1 &$1.170^{\star\star}$ &$0.522^{\star}$ &1.058 \\
			C2: KDE Margin \& InvCop2 &$1.197^{\star}$ &0.543 &1.087\\
			C3: Skew t Margin \& InvCop3 &$1.382^{\star\star\star}$ &$0.561^{\star}$ &1.088\\ \hline
			{\em State Space Models} & & & \\
			S1: SVUC &1.223 &0.538 &1.090\\
			S2: MSAR1 &1.241 &0.548 &1.097\\
			S3: UCAR4 &1.486 &0.574 &1.077\\  \hline \hline
		\end{tabular}
	\end{center}
	\caption{Summary of the accuracy of the one-step-ahead predictive
		distributions
		for the six models fit to the 
		U.S. electricity inflation data. The metrics are the mean (negative) logarithm predictive score (LP),
		the mean cumulative rank probability score (CRPS), and the 
		root mean squared error (RMSE). Lower values for all metrics indicate
		improved accuracy.
		The first three models employ inversion copulas
		constructed from latent state space models, along with an asymmetric margin. 
		The bottom three
		are the same state space models fit directly to the data. When the result
		for a copula model is statistically significantly better than the corresponding
		state space model, it is indicated with `$\star$' at 10\% significance level, `$\star\star$' at 5\% significance level or `$\star\star\star$' at 1\% significance level.}
	\label{WAtab:metrics}
\end{table}

%% file: figs_arxiv.tex
\begin{figure}[th]
\begin{center}
\includegraphics[width=380pt]{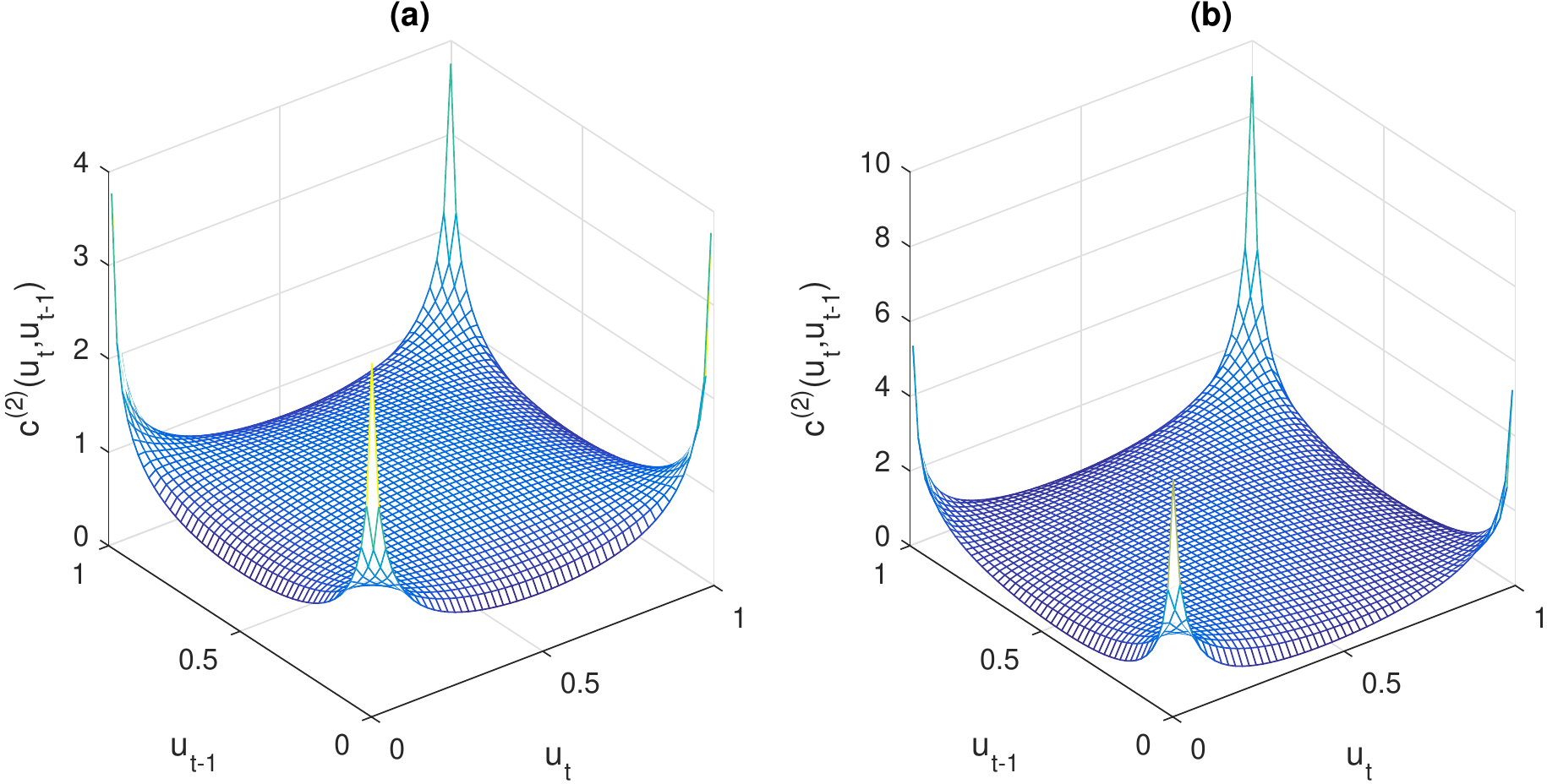}
\end{center}
\vspace{-15pt}
\caption{Bivariate marginal copula densities $c^{(2)}(u_t,u_{t-1}|\bm{\psi})$ 
of two copulas constructed by inversion of latent nonlinear 
state space models. Panel~(a) is for
a first order stochastic volatility model.
Here, the values of overall `level' dependence
(ie. Kendall's tau or Spearman's rho)
for this copula are exactly zero, yet the copula
has high (equally-valued) tail dependence in all four 
quadrants. Panel~(b) is for a Markov switching autoregression.
Here, dependence is asymmetric and quantile dependence
differs in each of the four quadrants.}
\label{fig:svcop}
\end{figure}

\begin{figure}[th]
\begin{center}
\includegraphics[width=380pt]{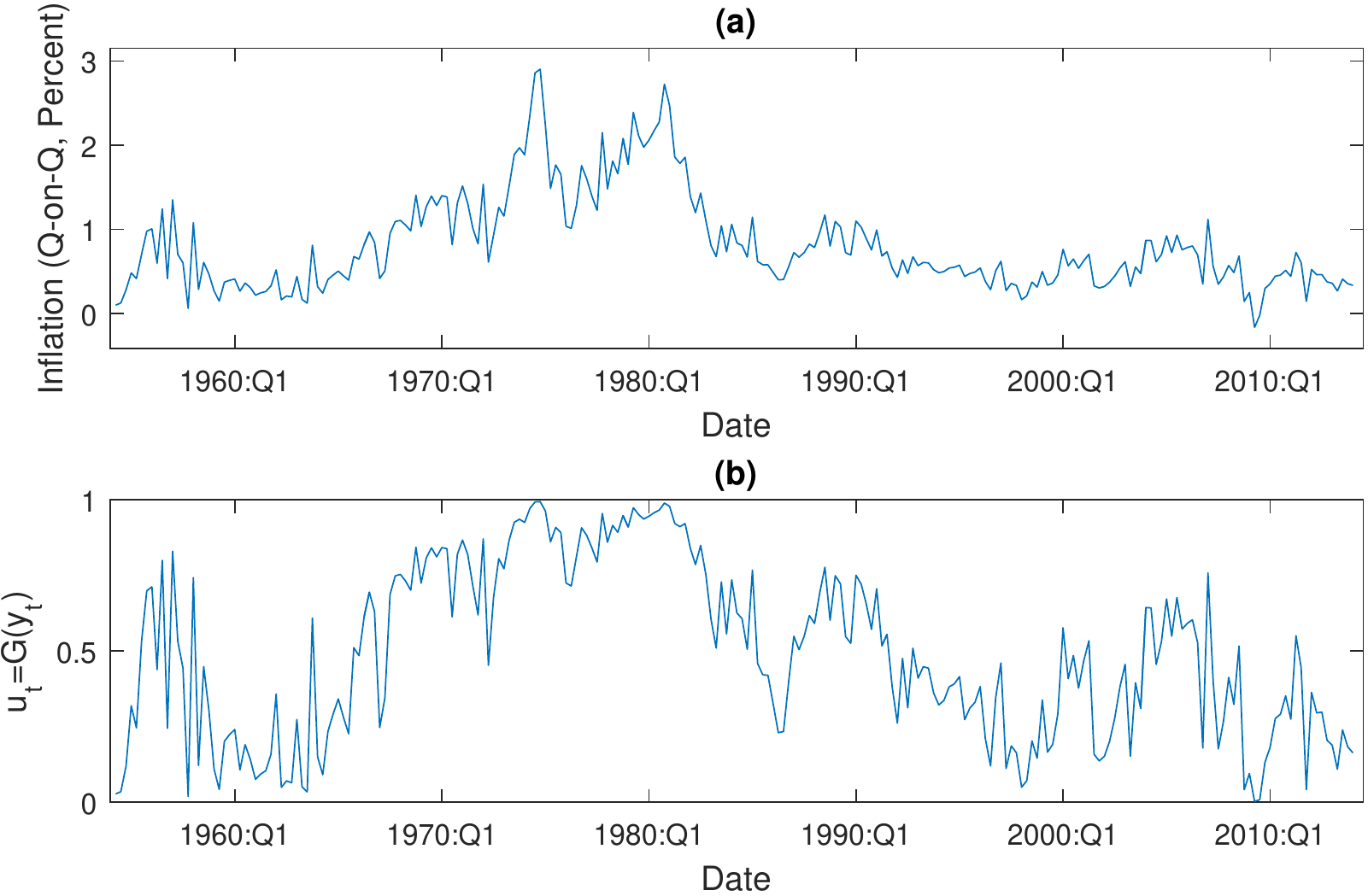}
\end{center}
\vspace{-15pt}
\caption{Panel~(a) is a time series plot
of the quarter-on-quarter U.S. broad inflation data.
Panel~(b) is a time series plot of the copula data $u_t=G(y_t)$,
where $G$ is the distribution function computed from the KDE in Figure~\ref{fig:margins}(a,b).}
\label{fig:infldata}
\end{figure}

\begin{figure}[h]
\begin{center}
\includegraphics[width=375pt]{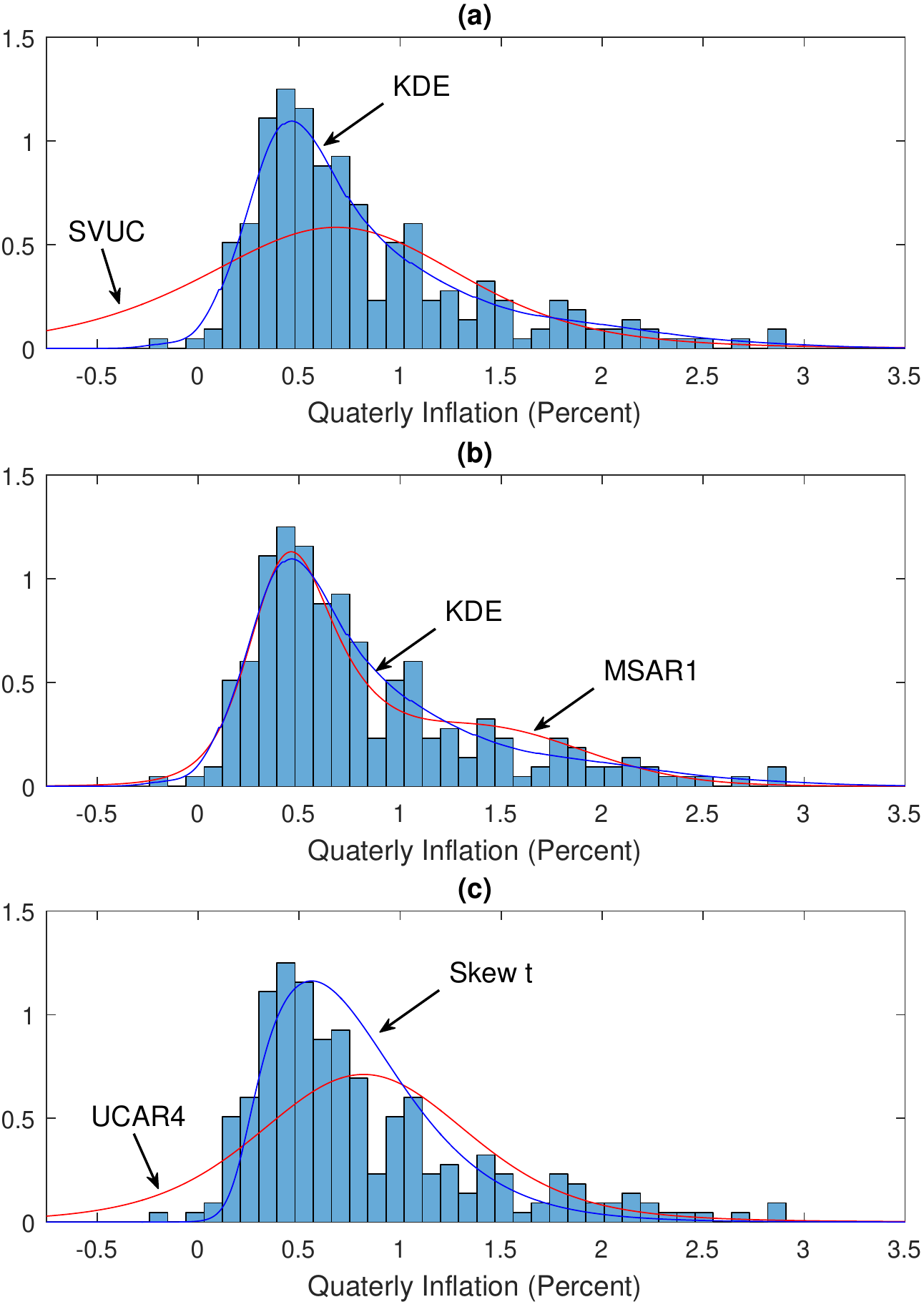}
\end{center}
\caption{Each panel plots the (normalized) histogram of the U.S. broad
inflation data, along with the marginal distributions of the six time series models.
Each panel plots the margin used for each of the three copula models in blue, along with
the margin arising from the corresponding state space model fit to the same data in red.}
\label{fig:margins}
\end{figure}

\begin{landscape}
	\begin{figure}[h]
		\centering
		\begin{tabular}{@{}ccc@{}}
			\includegraphics[width=170pt]{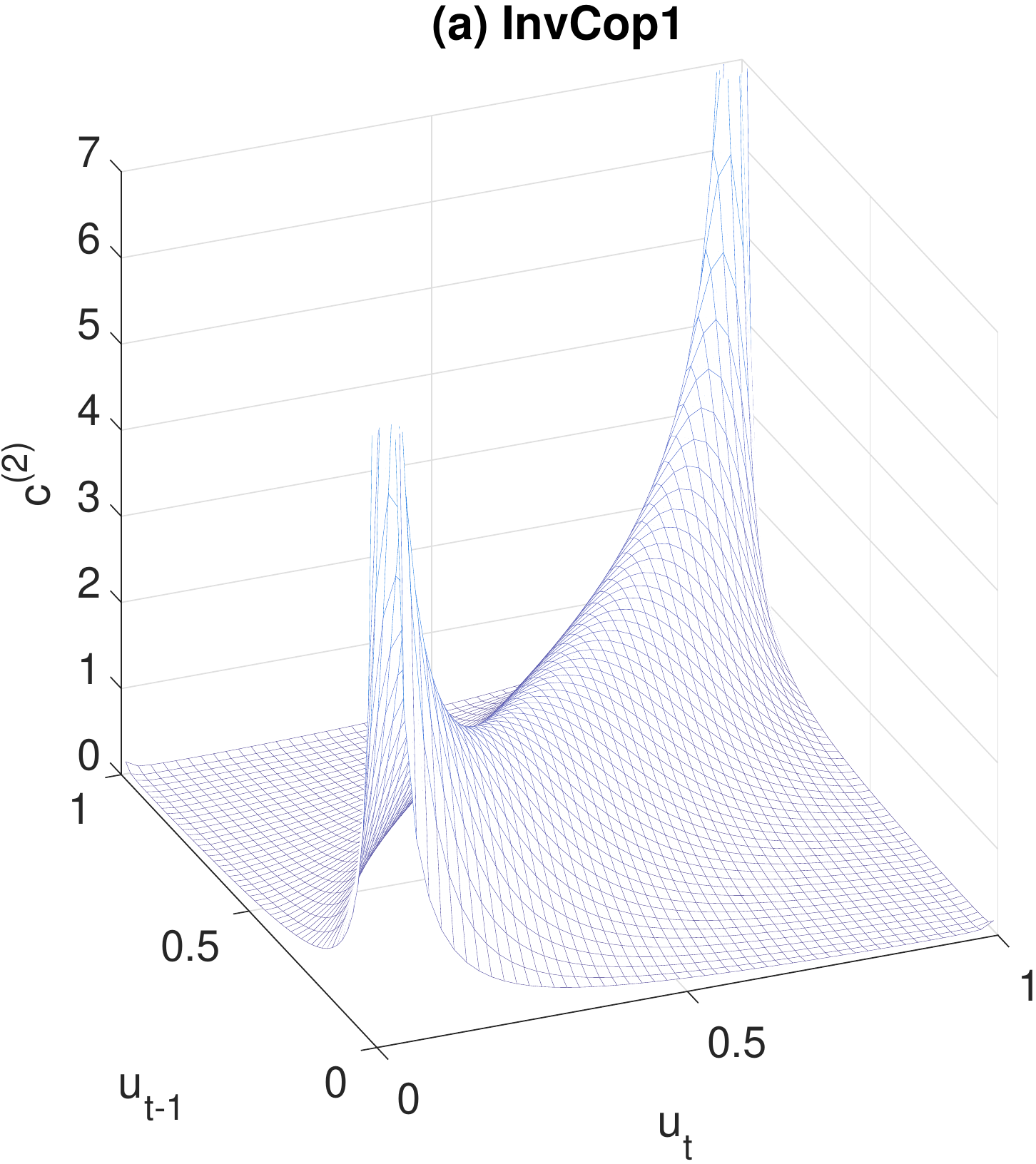} &
			\includegraphics[width=170pt]{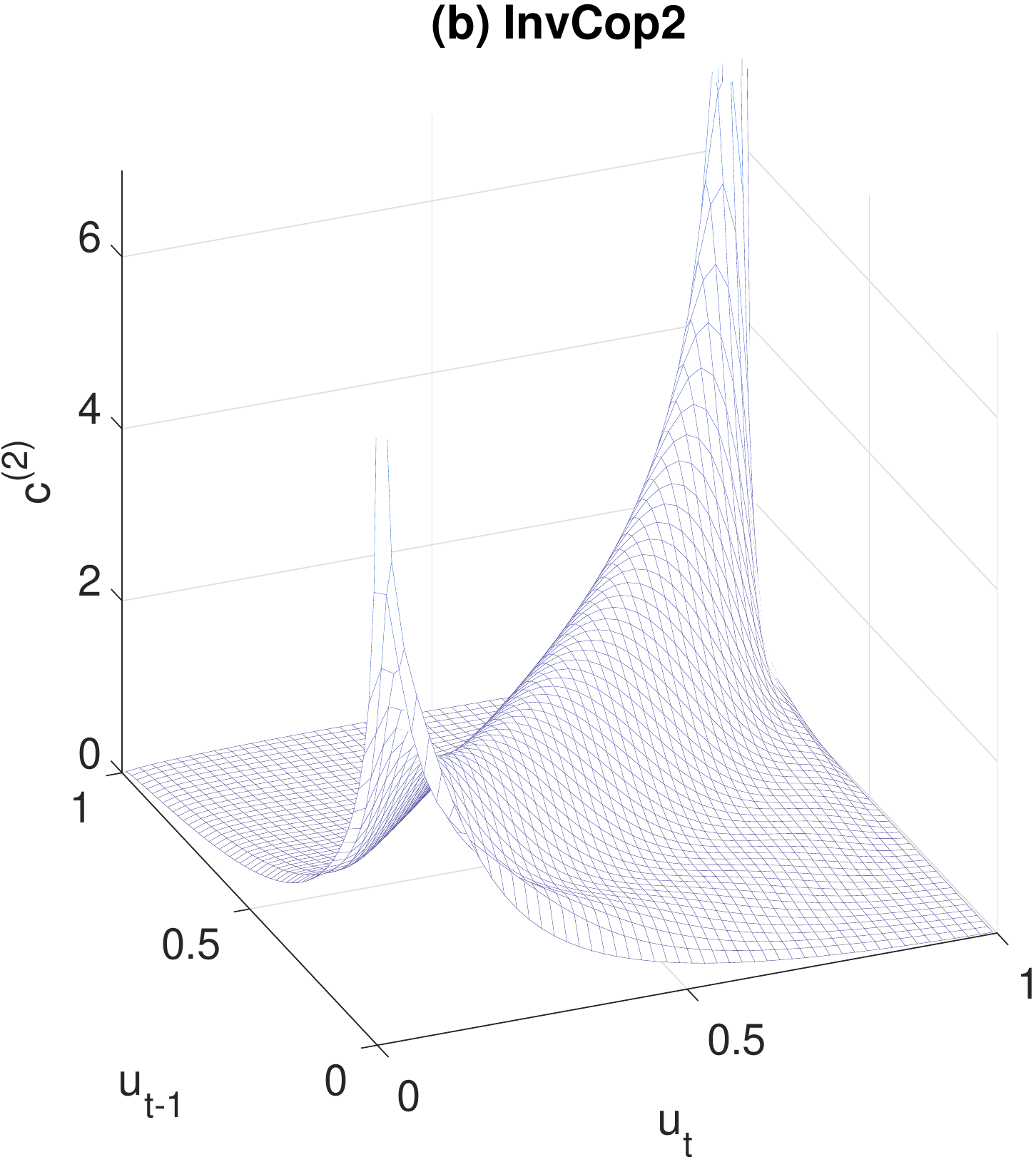} &
			\includegraphics[width=170pt]{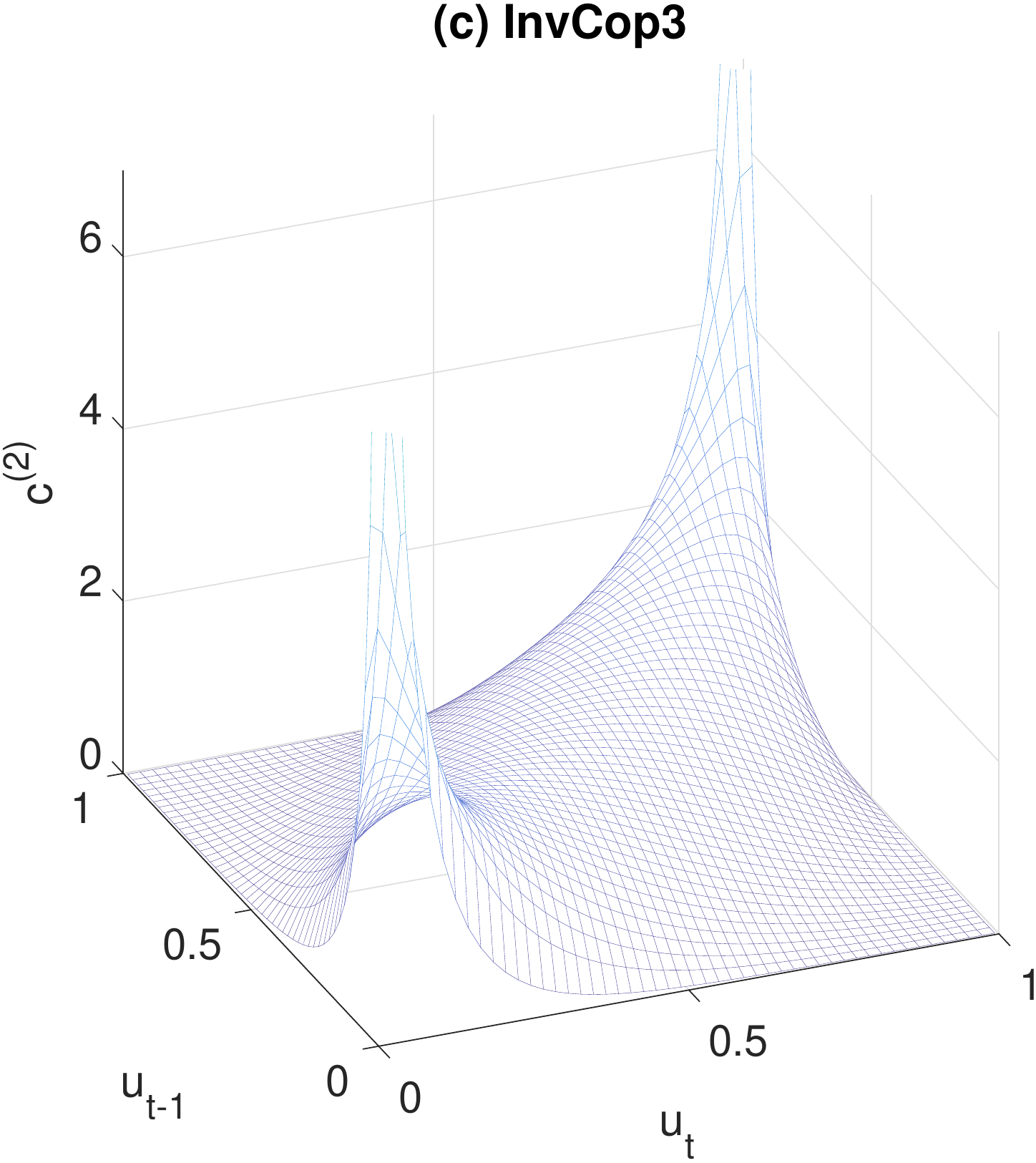} \\ \noalign{\vspace{15pt}}
			\includegraphics[width=170pt]{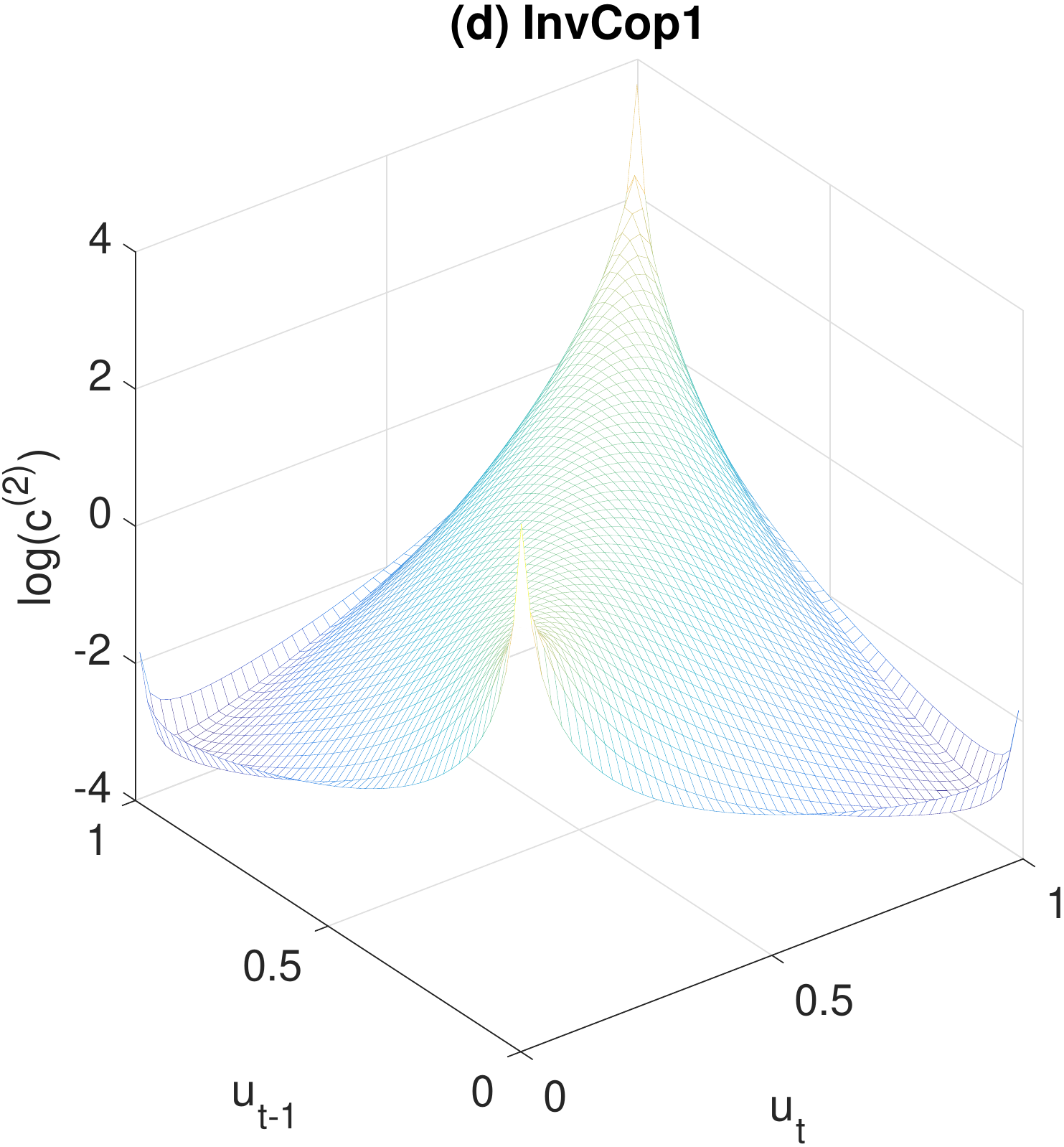} &
			\includegraphics[width=170pt]{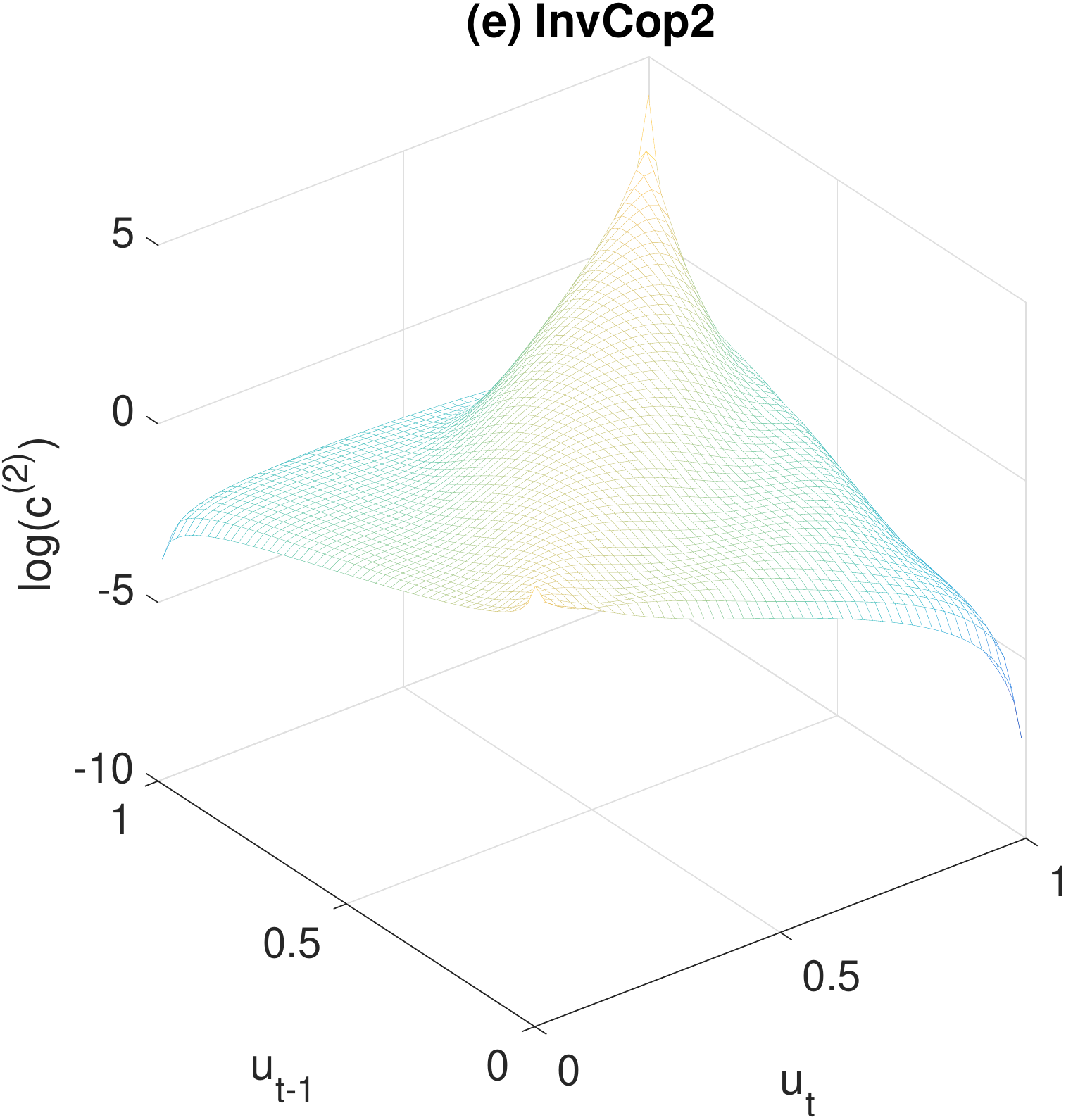} &
			\includegraphics[width=170pt]{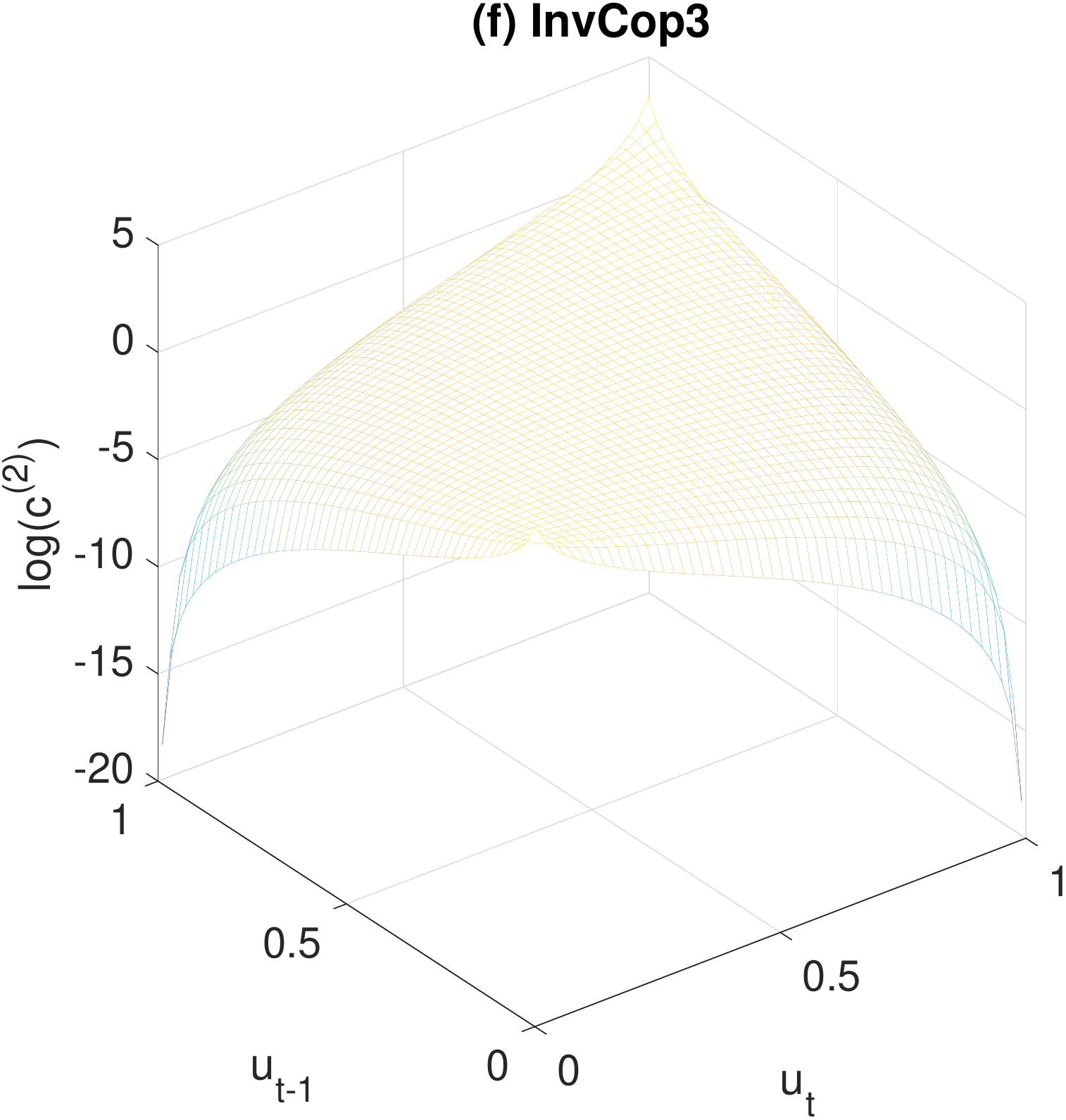} 
		\end{tabular}
		\caption{Marginal copula densities $c^{(2)}(u_t,u_{t-1}|\hat{\bm{\psi}})$
			for each of the 
			three inversion copula models fitted to the U.S. broad inflation data. 
			Panels~(a,b,c) plot the densities with a common
			vertical axis truncated at 7 for interpretation. 
			Panels~(d,e,f) plot the logarithm of the same three densities.
			Each density has been computed at the posterior mean $\hat{\bm{\psi}}$ of the
			copula parameters.}
		\label{fig:copdensity_comparison}
	\end{figure}
\end{landscape}

\begin{landscape}
\begin{figure}[h]
\begin{center}
\includegraphics[width=600pt]{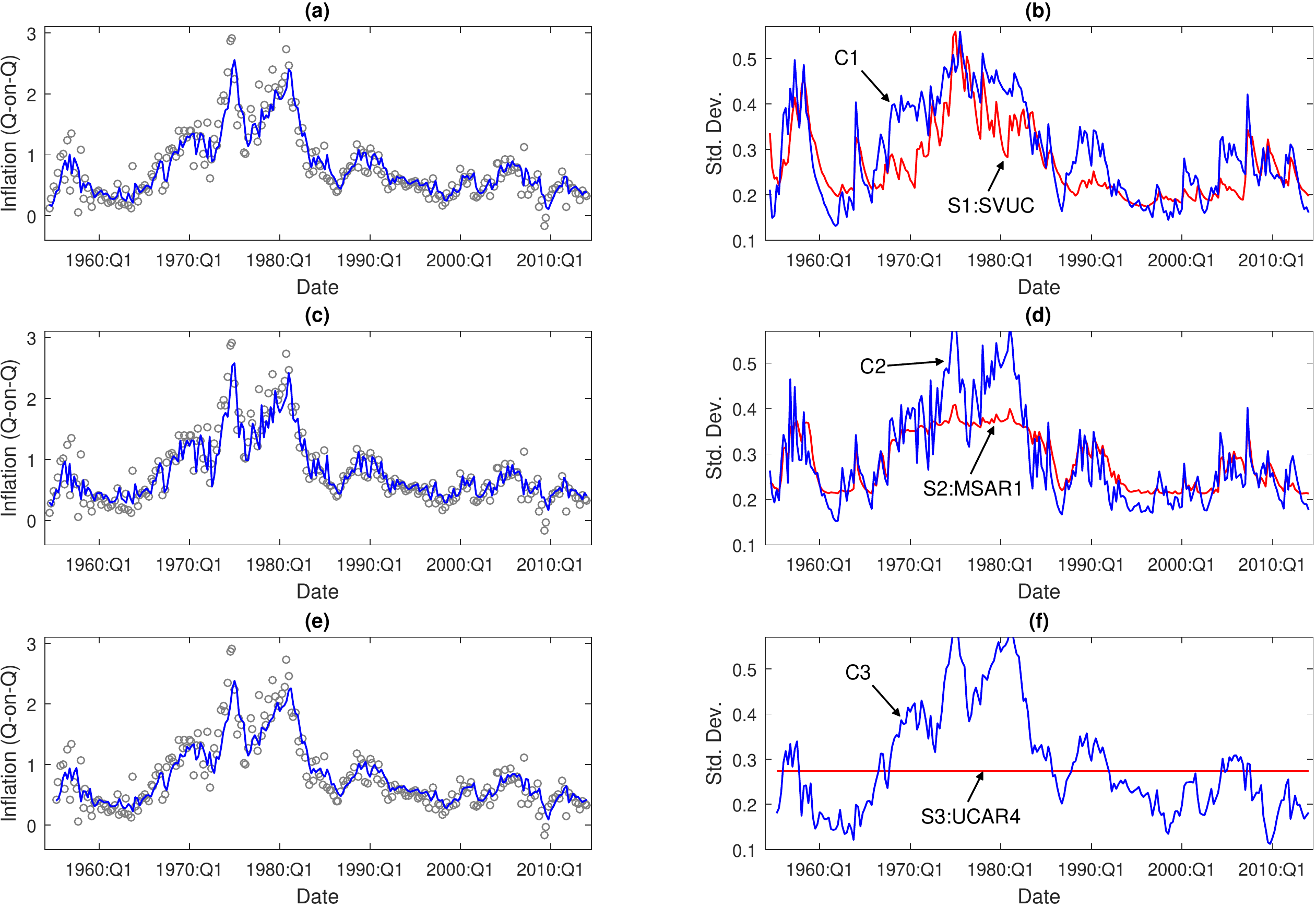} 
\end{center}
\caption{Moments of the one-step-ahead predicitve distributions of the U.S. broad inflation data. Panels~(a,c,e) plot in blue the 
predictive means from each of the three copula models C1, C2 and C3, respectively (See 
Table~\ref{tab:mmetric} for their specification). Also plotted is
a scatterplot of the data. Panels~(b,d,f) plot the standard deviations of the 
predictive distributions
from the three copula models in blue. The standard deviations of the predictive 
distributions from the three
corresponding state space models, fit directly to the same
data, are plotted in red.}
\label{fig:pred_moments}
\end{figure}
\end{landscape}

\begin{figure}[h]
\begin{center}
\includegraphics[width=380pt]{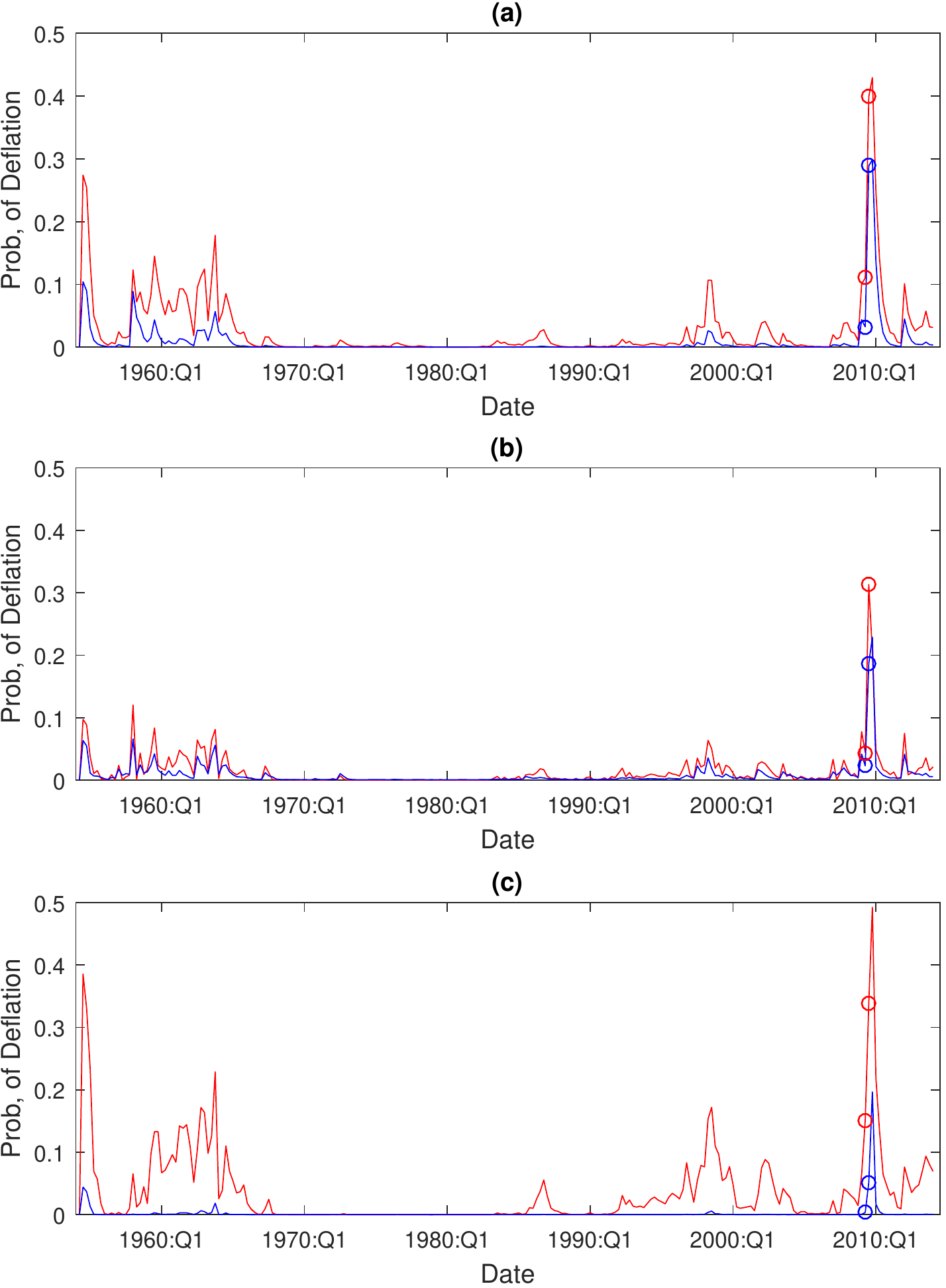} 
\end{center}
\caption{The one-step-ahead predictive probability of deflation from each of 
the six models fit to the U.S. broad inflation data. The probabilities arising from
the copula models are in blue, and the state space models fit directly to the data
in red.
Panel~(a) plots these for models C1 and S1; 
panel~(b) for models C2 and S2; and panel~(c) for models
C3 and S3. Circles denote the
two quarters where (very mild) deflation is recorded in our data
(-0.167\% during 2009:Q1, and -0.022\% during 2009:Q2).}
\label{fig:deflation}
\end{figure}

\begin{figure}[h]
	\begin{center}
		\includegraphics[width=440pt]{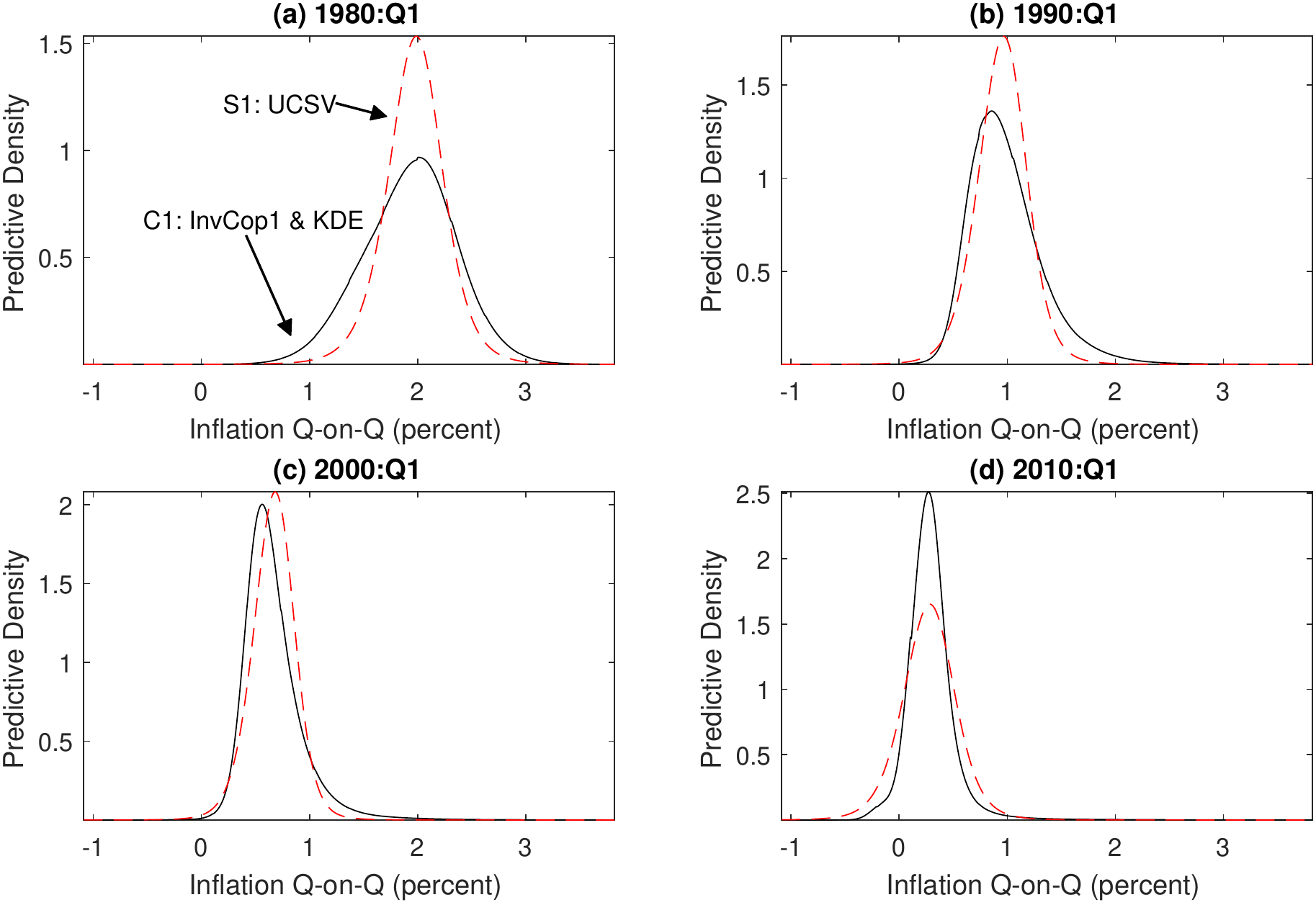} 
	\end{center}
	\caption{One-quarter-ahead predictive densities of U.S. broad inflation from the UCSV state space model
		fit directly to the data (S1; red dashed line), and for the copula time series model with
		copula InvCop1 and KDE margins (C1; black solid line). Results are presented for four quarters: (a) 1980:Q1, (b) 1990:Q1, (c) 2000:Q1, (d) 2010:Q1. 
		Note that the densities from C1 exhibit different skew and kurtosis.}
	\label{fig:4dens}
\end{figure}  

\begin{figure}[htbp]
	\begin{center}
		\includegraphics[width=320pt]{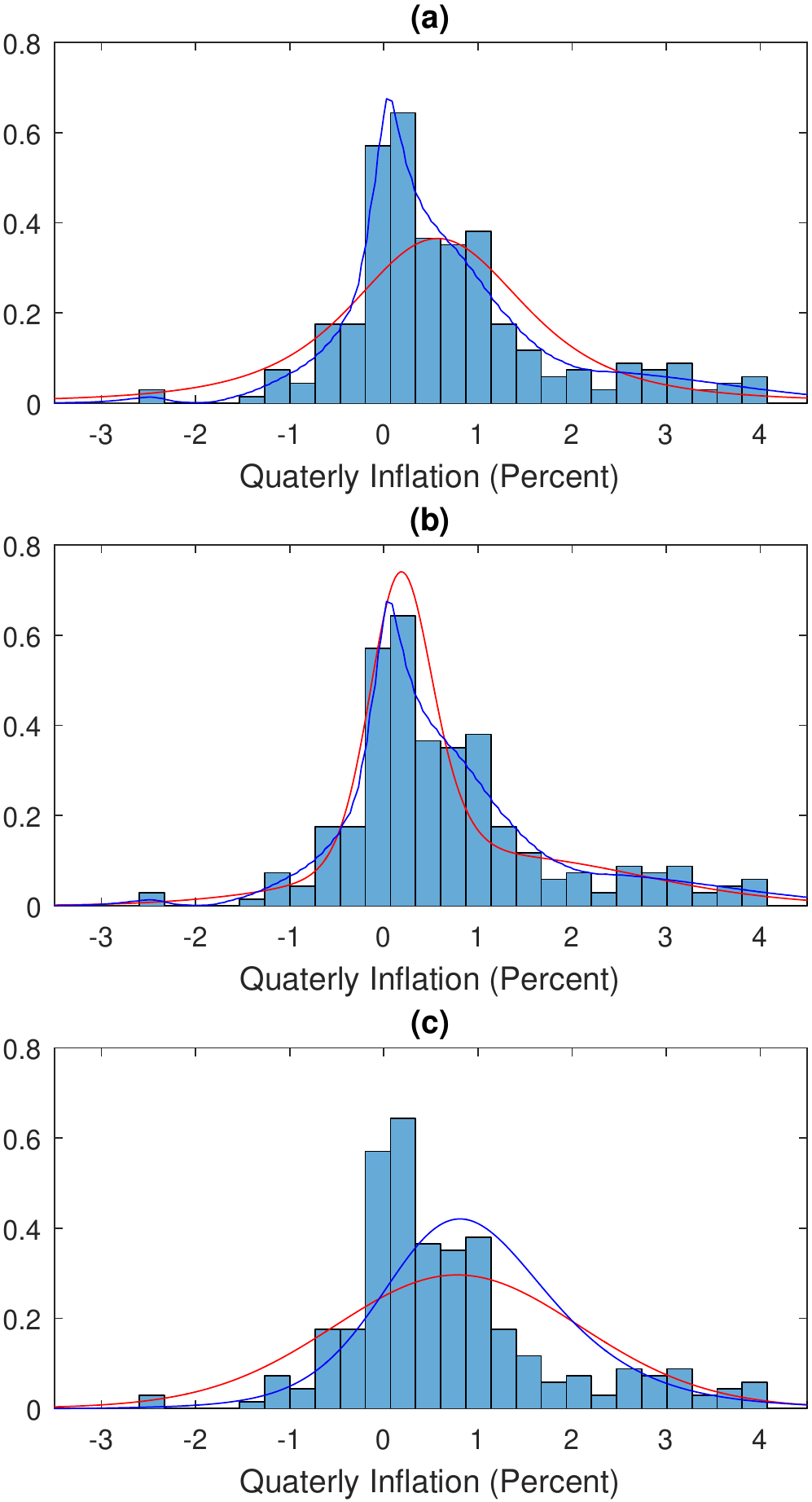} 
	\end{center}
	\caption{Each panel plots the (normalized) histogram of the U.S. 
		electricity
		inflation data, along with the marginal distributions of the six time series models.
		Each panel plots the margin used for each of the three copula models
		in blue, along with
		the margin arising from the corresponding state space model fit to the same data in red.}
	\label{WAfig:margins}
\end{figure}